\def\astroph{1}
  \ifnum\astroph=0
     \documentclass[manuscript]{aastex}          
  \else
    \documentclass{emulateapj}
   \fi

  \usepackage{epsfig,amsmath,natbib}
  \usepackage[dvips]{color}
  \usepackage{verbatim}

\begin{document}
 \slugcomment{2nd draft}
\shorttitle{Ly$\alpha$ RT with dust:
            escape fractions from simulated high-$z$ galaxies}
\shortauthors{Laursen et al.}
\title{Lyman $\alpha$ Radiative Transfer with Dust:\\
       Escape Fractions from Simulated High-Redshift Galaxies}
\author{Peter Laursen\altaffilmark{1},
        Jesper Sommer-Larsen\altaffilmark{3,1}
        and Anja C. Andersen\altaffilmark{1}}
%
%
\altaffiltext{1}{Dark Cosmology Centre, Niels Bohr Institute, University of
                 Copenhagen, Juliane Maries Vej~30, DK-2100, Copenhagen {\O},
                 Denmark; email: pela@dark-cosmology.dk}
\altaffiltext{3}{Excellence Cluster Universe, Technische Universit\"at
                 M\"unchen, Boltzmannstra\ss e 2, D-85748 Garching,
                 Germany; email:\\ jslarsen@astro.ku.dk}

\begin{abstract}
The Lyman $\alpha$ emission line is an essential diagnostic tool for probing
galaxy formation and evolution. Not only is it commonly the strongest
observable line from high-redshift galaxies
but from its shape detailed information about its host galaxy can be revealed.
However, due to the scattering nature of Ly$\alpha$ photons increasing their
path length in a non-trivial way, if dust is present in the galaxy the line may
be severely suppressed and its shape altered.
In order to interpret observations correctly, it is thus of crucial
significance to know how much of the emitted light actually escapes the galaxy.

In the present work,
using a combination of high-resolution cosmological
hydro-simulations and an adaptively refinable Monte Carlo Ly$\alpha$ radiative
transfer code including an advanced model of dust, the escape fractions
$f_{\mathrm{esc}}$ of Ly$\alpha$ radiation
from high-redshift ($z = 3.6$) galaxies are calculated.
In addition to the average escape fraction, the variation of
$f_{\mathrm{esc}}$ in different directions and from different parts of the
galaxies is investigated, as well as the effect on the emergent
spectrum.

Escape fractions from a sample of simulated galaxies of representative
physical properties are found to decrease for increasing galaxy virial mass
$M_{\mathrm{vir}}$, from $f_{\mathrm{esc}}$ approaching unity for
$M_{\mathrm{vir}} \sim 10^9$ $M_\odot$ to $f_{\mathrm{esc}}$ less than 10\% for
$M_{\mathrm{vir}} \sim 10^{12}$ $M_\odot$.
In spite of the dust being
almost grey, it is found that the emergent spectrum is affected non-uniformly,
with the escape fraction of photons close to the line center being much higher
than of those in the wings, thus effectively narrowing the Ly$\alpha$ line.
\end{abstract}


\section{Introduction}
Many astrophysical and cosmological key questions depend upon precise
measurements of the luminosities of distant galaxies;
in particular, galactic star
formation rates (SFRs) and histories,
as well as luminosity functions (LFs),
are crucially contingent on the amount
of assumed luminosities.
A fundamental problem in this context is naturally the question of how large a
fraction of the emitted light actually escapes the galaxy.
If an unknown fraction of the emitted light is absorbed, either by gas or by
dust, the inferred quantity of interest clearly will be subject to large
uncertainties or, at best, a lower limit.

The Ly$\alpha$ line, albeit notoriously challenging to interpret, is an
extremely powerful probe of the high-redshift Universe.
At $z > 2.1$, the cosmological redshift of Ly$\alpha$ makes it the
strongest emission line in the optical-NIR window, 
while for $z \gtrsim 4$, Ly$\alpha$ emitters (LAEs; galaxies detected from
their emission
in Ly$\alpha$, either by narrow-band photometry or spectroscopy) becomes
easier to
detect than Lyman-break galaxies (LBGs; galaxies detected from their dropout
in wavelengths blueward of the Lyman-break).

From the shape of the Ly$\alpha$ line profile, a variety of valuable
information can be obtained about the galaxies emitting it. Ly$\alpha$
resonant scattering theory as well as simulations tell us that, in general,
due to the large optical depth of the neutral hydrogen for a line center
photon, the radiation must diffuse to either the blue or the red side of the
center, and consequently should escape the interstellar medium (ISM)
of its host galaxy in a
double-peaked profile. The exact shape of this profile depends on the physical
state of the gas, such as its column density, temperature, velocity field, and,
in particular, dust contents. Similarly, the strength of the line depends on
the stellar population and the dust contents. Notwithstanding the complexity,
and indeed sometimes intertwinement, of these dependencies, this also makes
Ly$\alpha$ a potentially strong source of information.

Whereas absorption processes in gas in many cases are well-known,
the effect of dust on the radiative transfer (RT) is still an intensely debated
subject. In contrast to gas, laboratory experiments with dust are extremely
complex, partly due to the complications involved in replicating the physical
environments of the ISM, partly due to our limited
knowledge vis-\`a-vis what actually constitutes cosmic dust.

In the present-day Universe, most dust is formed in the atmospheres of stars on
the asymptotic giant branch (AGB) of the Hertzsprung-Russel diagram; the dying
phase of stars less massive than $\sim$8 $M_\odot$ \citep[e.g.][]{hof07,mat08}.
In these environments the gas is sufficiently cool, yet sufficiently dense
that molecules may form and stick together to form dust grains.

However, there is observational evidence that dust is also abundantly present
in the early Universe \citep[e.g.][]{str07,cop09}.
Since the time to reach
the AGB phase is of the order of 1 Gyr,
something else must have provided the
ISM with dust at these epochs. A promising candidate is supernovae (SNe),
the ejecta of which are believed to exhibit favorable conditions
for the formation of dust for a short period of time, approximately 600 days
after the explosion \citep[e.g.][]{kot09}.

For the Milky Way (MW), as well as  the Small and Large Magellanic Clouds
(SMC; LMC),
the dust extinction curves, i.e.~the extinction of light at a given wavelength,
are fairly well established \citep[e.g.][]{bia96,nan81,pre84,pei92}, and from
the observed color excess $E(B-V)$ one may then derive the total extinction.
The term ``extinction'' refers to removal of light from the line of sight,
be it due to absorption or scattering, and may be characterized by the number
$A_\lambda$ of magnitudes by which the observed object is diminished.
For more distant galaxies one is usually obliged to assume similar extinction
curves. Since the stellar population of
the SMC is younger than that of the LMC, an SMC extinction curve might be
expected be able to describe better the dust in high-redshift galaxies, and
has indeed proved to be a good fit in GRB host galaxies \citep[e.g.][]{jak04}.
Note, however, that the prominent feature at 2175 {\AA}, characteristic of the
LMC and MW extinction, has been detected in a few cases also at high redshift
\citep{jun04,ell06,sri08,eli09}.

The overall normalization of extinction curves comes from the observed
property that the extinction is found to be very close to proportional with the
column density $N_{\mathrm{H}}$ of hydrogen \citep[e.g.][]{boh78}.
Typically, one combines measurements of 
$N_{\textrm{{\scriptsize H}{\tiny \hspace{.1mm}I}}}$
(and $N_{\textrm{{\scriptsize H}}_2}$)
with the extinction in the $V$-band, $A_V$. In this way, one then knows how
much light is extinguished when traveling a given physical distance in space.

However, for light that does not travel directly from the source to the
observer, as is the case for resonantly scattered lines like Ly$\alpha$, the
situation becomes more raveled. Not only does the total distance covered by the
photons increase by a large and a priori unknown factor, but the photons
received from a given point on the sky may also have traveled through
physically different environments, in turn implying an unknown
and possibly highly increased probability of being absorbed by dust.

For this reason, the observed fact that Ly$\alpha$ radiation nonetheless
\emph{does}
escape has long puzzled astronomers. The fact that Ly$\alpha$ line profiles
are often seen to exhibit a P Cygni-like profile has led to the suggestion
that high-velocity outflows of gas are needed to enable escape
\citep{kun98,ost08,ate08}. However, at high redshifts many galaxies are still
accreting matter, which should result in an increased \emph{blue} peak.
Since this is rarely observed, the shape could be caused by other mechanisms,
e.g.~IGM absorption.

The angle under which a galaxy is viewed may also affect the amount of
observed radiation. Ionizing UV radiation could create ``cones'' of low neutral
hydrogen density emanating from the star-forming regions through which the
Ly$\alpha$
can escape \citep{ten99,mas03}. Even without these ionized cones, scattering
effects alone may cause an anisotropic escape of the Ly$\alpha$ \citep{lau07}.

Another commonly repeated scenario is a
multi-phase medium, where the dust is locked up in cold clouds so that the
photons primarily travel in an ionized, dustless medium \citep{neu91,han06}.
Since continuum radiation travels through the cloud, it would be attenuated
more by the dust. This could explain the high Ly$\alpha$ equivalent widths
occasionally observed in in LAEs \citep[e.g.][]{mal02,rho03,shi06}.

Previous attempts to determine Ly$\alpha$ escape fractions from high-redshift
galaxies have mainly been
trying to match observed Ly$\alpha$ luminosities with expected, and different
methods obtain quite different results.
\citet{led05,led06} found very good agreement between galaxies simulated with
the galaxy formation model {\sc galform} and observational data at
$z = 3$--6, using a constant escape fraction of $f_{\mathrm{esc}} = 0.02$ and
assuming no IGM absorption.
\citet{dav06} obtained similar results by matching the Ly$\alpha$ LF
of galaxies from their cosmological SPH simulation to the data of
\citet{san04}, although \citet{nag08} argued that the data is based on a small
sample and that the simulation box size is too small. Matching the simulated
Ly$\alpha$ LF to the observed one by \citet{ouc08}, \citet{nag08} themselves
obtain $f_{\mathrm{esc}} \simeq 0.1$, although the preferred scenario is not
that a certain fraction of the Ly$\alpha$ radiation escapes, but rather that
a certain fraction of LAEs are ``turned on'' at a given time (the so-called
``duty cycle scenario'').
In a similar way, \citet{day09} find somewhat higher escape fractions at
$z\sim5.7$ and $\sim$6.5 ($f_{\mathrm{esc}} \sim 0.3$), which they use for
predicting the LF of LAEs at $z\sim7.6$.

\citet{gro07} compared inferred Ly$\alpha$ and rest-frame UV continuum SFRs of
a large sample of LAEs from the MUSYC \citep{gaw06a} survey and argue that an
escape fraction of $\sim$$1/3$ is needed to explain the discrepancy,
although \citet{nil09} pointed out that a missing $(1+z)$-factor probably
explains the difference.
Matching  Ly$\alpha$-inferred SFRs to SED modeling of
observed LAEs, \citet{gaw06b} found an $f_{\mathrm{esc}}$ of $\sim$0.8, with a
lower limit of 0.2. While SED fitting may not be the most accurate way of
estimating SFRs, aiming to match these observations, \citet{kob07} obtain
similar result theoretically by incorporating the effects of galactic outflows.

In this paper we aim to scrutinize the effect of dust on the Ly$\alpha$ RT
by applying the Monte Carlo RT code {\sc MoCaLaTA} \citep{lau09} on a number of
simulated galaxies, extracted from fully cosmological TreeSPH simulations.
The theoretical aspects of cosmic dust is discussed in
\S\ref{sec:theo} and the implementation of this into the RT code in
\S\ref{sec:imp}. The new version of the code is then tested against physically
idealized situations for which analytical solutions exist in \S\ref{sec:test},
before it is applied
to realistic simulated galaxies in \S\ref{sec:sim}. The results, and the
sensitivity of these on the values of different input parameters, are presented
in \S\ref{sec:res} and \S\ref{sec:parstud}, respectively. Finally, the
obtained results are summarized and discussed in \S\ref{sec:sum}.


\section{Theory of Dust}
\label{sec:theo}

Four quantities characterize what impact the dust grains will have on the
propagating Ly$\alpha$ photons:
the \emph{density};
the (wavelength dependent) \emph{cross-section} of interaction;
the \emph{albedo} giving the probability that a photon incident on a dust
grain will be scattered rather than absorbed;
and finally the \emph{phase function} defining the direction
into which a non-absorbed photon is scattered.
These quantities will be discussed below.

Dust grains are built up from metals, and thus the dust density is
expected to scale with gas metallicity in some fashion.
Metals are created in
dying stars, i.e.~in AGB stars and SNe. For sufficiently dense and cold
environments, the neutral metals form molecules which eventually stick together
to form dust. No formal definition of the distinction between large molecules
and dust grains exist, but may be taken to be of the order of $\sim$500 atoms
or so.

Depending on the abundances of the individual metals, as well as the physical
conditions, a variety of different types of dust may be produced, with
regards to both composition and structure, and hence with different scattering
properties. Much effort has been put into unraveling the nature of cosmic dust,
in particular in explaining the 2175 {\AA} bump. This feature is generally
attributed to carbonaceous materials, e.g.~graphite, diamonds, and/or
polycyclic aromatic hydrocarbons, but still the precise nature remains unknown.

In principle, the result of a photon interacting with a dust grain may be
calculated analytically by solving Maxwell's equations, on the basis of the
geometry of the particle and its optical properties, i.e.~the dielectric
functions. This is trivial in the case of simple geometries such as spheres and
spheroidals. More general shapes and composites can be modeled by discretizing
the grain into a large number of dipoles; the so-called \emph{discrete dipole
approximation} \citep{pur73,dra88}, but for the complex and, more importantly,
uncertain or unknown shape of realistic grains, this is not feasible.

Had we full knowledge of the relevant properties of dust, a distribution of the
various species could be calculated in
simulated galaxies, and the radiative transfer could then be realized by
computing the total optical depth of the ISM as a sum
of all contributors, and determining for each scattering the kind of particle
responsible for the scattering.
However, lacking a sound theory of the formation of dust grains, in particular
in the high-redshift Universe, we take a
different approach: although the exact nature of cosmic dust is not known,
the average extinction --- and hence the cross-sectional area --- of dust as a
function of wavelength is known for many different sightlines through the SMC
and the LMC \citep[e.g.][]{gor03}.
Since the metallicity of the Magellanic Clouds is fairly well
known, the extinction curve of the SMC (or LMC) can be scaled to the
metallicity of the simulated galaxies, thus yielding
the extinction in the simulations.

\subsection{Cross-section}
\label{sec:xsecd}

Observationally, the extinction $A_V$ in the $V$-band is found to have a
surprisingly constant proportionality with the column density of hydrogen from
sightline to sightline within the MW \citep[e.g.][]{mas86,fit07}.
Similar results, but with
different normalizations, are found for the SMC and the LMC \citep{gor03}.
Accordingly, the cross-section $\sigma_{\mathrm{d}}(\lambda)$ of dust may be
conveniently expressed as an effective cross-section \emph{per hydrogen atom},
thus eliminating any assumptions about the size
distribution, shape, etc., and merely relying on observed extinction curves.
The optical depth $\tau_{\mathrm{d}}$ of dust when traveling a distance $r$
through a region of hydrogen density $n_{\mathrm{H}}$ is then
\begin{equation}
\label{eq:taud}
\tau_{\mathrm{d}} = n_{\textrm{{\scriptsize H}}} r \sigma_{\mathrm{d}}
                  = N_{\textrm{{\scriptsize H}}}   \sigma_{\mathrm{d}}.
\end{equation}

The quantity usually measured is $A_\lambda/N_{\mathrm{H}}$,
and the cross-section is then 
\begin{equation}
\label{eq:sigdth}
\sigma_{\mathrm{d}} = \frac{\ln 10}{2.5} \frac{A_\lambda}{N_{\mathrm{H}}}
\end{equation}
We use the fit to the SMC or LMC extinction curves proposed by \citet{pei92},
which is an extension of the \citet{mat77}-model. The fit is a sum of six terms
(Drude profiles) representing a background, a
far-ultraviolet (FUV), and a far-infrared (FIR) extinction, as well as the
2175 {\AA}, the 9.7 $\mu$m, and the 18 $\mu$m extinction features.
Based on newer data from \citet{wei01}, \citet{gne08}
adjusted the fit and added a seventh term to account for the narrow, asymmetric
FUV peak in the dust extinction.

\begin{figure}
\epsscale{1}
\plotone{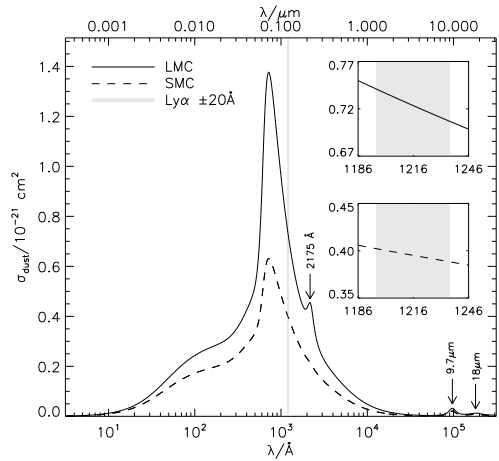}
\caption{Extinction cross-section fits to the observed extinction curves of the
         LMC (\emph{solid}) and the SMC (\emph{dashed}).
         The difference in amplitude is mainly due to the SMC being less
         metal-rich than the LMC. The vertical, grey-shaded area is the region
         inside which the (rest-frame) Ly$\alpha$ line is expected to fall.
         The two inlets show a zoom-in of this region on the extinction curves
         (\emph{top}: LMC, \emph{bottom}: SMC), demonstrating the linearity
         across the Ly$\alpha$ line.}
\label{fig:gnedin}
\end{figure}
Figure \ref{fig:gnedin} shows these fits. The inlets show that the extinction
curves are very close to being linear in the vicinity of the Ly$\alpha$ line.
In fact, in this region it is an excellent approximation to write the
cross-section as

 

\begin{eqnarray}
\label{eq:sigd}
\nonumber
\sigma_{\mathrm{d}}/10^{-21}\textrm{ cm}^2 = & & \\
& \hspace{-2.5cm}\left\{ \begin{array}{ll}
0.395 + 1.82\times10^{-5}\,(T/10^4\textrm{ K})^{1/2}\, x & \textrm{ for the SMC}\\
0.723 + 4.46\times10^{-5}\,(T/10^4\textrm{ K})^{1/2}\, x & \textrm{ for the LMC}.
\end{array} 
\right. &
\end{eqnarray}

Here, we have used the parametrization
$x \equiv (\nu-\nu_0)/\Delta\nu_{\mathrm{D}}$ of the frequency $\nu$ of the
photon, where $\nu_0$ is the line center frequency and
$\Delta\nu_{\mathrm{D}}$
is the Doppler width of the line (times $\sqrt{2}$) resulting from the thermal
motion of the atoms with temperature $T$. Note that $T$ only enters
Eq.~(\ref{eq:sigd}) to account for the temperature dependency of $x$;
$\sigma_{\mathrm{d}}$ itself is independent of $T$.


\subsection{Number density}
\label{sec:dens}

The reason for the variability of extinction with galaxy, and the
non-variability with sightline, is to a large degree the different overall
metallicities of the galaxies.
Although differences exist within the galaxies, the differences are larger from
galaxy to galaxy \citep{pei92}. In most of the calculations we will use an SMC
curve, but as shown in \S\ref{sec:parstud} the result is not very different if
an LMC curve is used.

Because the cross-section is expressed as a cross-section per hydrogen atom,
the relevant quantity is not dust density, but hydrogen density.
However, since in general the metallicity at a given location in a
simulated galaxy differs
from that of the Magellanic Clouds, the amplitude of the extinction will also
differ. Assuming that extinction scales with metallicity, a corresponding
pseudo number density $n_{\mathrm{d}}$ of dust at a given location of hydrogen
density $n_{\textrm{H}}$ and metallicity $Z_i$ of element $i$ can then be
calculated as
\begin{equation}
\label{eq:ndpre}
n_{\mathrm{d}} \sim n_{\textrm{H}} \frac{\sum_i Z_i}{\sum_i Z_{i,0}},
\end{equation}
where $Z_{i,0}$ is the average metallicity of element $i$ in the galaxy
the extinction curve of which is applied.
Obviously, $n_{\mathrm{d}}$ is not a true dust number density, but merely a
rescaled hydrogen number density.

On average, the SMC metallicities of the different elements are deficit
relative to Solar values by 0.6
dex \citep[e.g.][]{wel97}, while the LMC is deficit by 0.3 dex
\citep[e.g.][]{wel99}.
Small metal-to-metal deviations from this exist, but scaling $Z_i$ to the
metallicity of the
individual metals, using values from \citet{rus92}, does not change the
outcome significantly (cf.~\S\ref{sec:parstud}).

The reason that Eq.~(\ref{eq:ndpre}) is not expressed as a strict equality is
that we have so far neglected to differentiate between neutral and ionized
hydrogen.
Dust grains may be destroyed in a number of ways, e.g.~through collisions with
other grains,
sputtering due to collisions with ions, sublimation or evaporation, or even
explosions due to ultraviolet radiation
\citep[e.g.][]{gre76}.
These scenarios are all expected to become increasingly important for
hotter environments.
Accordingly, studies of the interstellar abundances of dust have usually
assumed that ionized regions contribute negligibly to the dust density,
and merely concerned themselves with measuring
column densities of neutral hydrogen, i.e.~H{\sc i} + H$_2$.
Moreover, many metallicity measurements are derived from low-resolution spectra
not capable of resolving and characterizing various components of the ISM.
As discussed in \S\ref{sec:iongas}, dust is also observed in regions
that are primarily ionized, and since the bulk of the Ly$\alpha$ photons is
produced in the proximity of hot stars with a large intensity of ionizing UV
radiation, even a little dust associated with the ionized gas might affect the
results.

Hence, we assume that the amount of dust scales with the total amount of
neutral hydrogen \emph{plus} some fraction
$f_{\mathrm{ion}}$ of the ionized hydrogen,
and Eq.~(\ref{eq:ndpre}) should then be
\begin{equation}
\label{eq:nd}
n_{\mathrm{d}} = (n_{\textrm{{\scriptsize H}{\tiny \hspace{.1mm}I}}}
               +  f_{\mathrm{ion}}
                  n_{\textrm{{\scriptsize H}{\tiny \hspace{.1mm}II}}})
                  \frac{\sum_i Z_i}{\sum_i Z_{i,0}} 
\end{equation}

Of course, this is not a physical number density of dust grains but with this
expression, the total optical depth of gas and dust as seen by a propagating
photon traveling a distance $r$ is
\begin{equation}
\label{eq:Ntot}
\tau_{\mathrm{tot}} = r (n_{\textrm{{\scriptsize H}{\tiny \hspace{.1mm}I}}}
                         \sigma_x
                    + n_{\mathrm{d}} \sigma_{\mathrm{d}}),
\end{equation}
where the neutral hydrogen cross-section $\sigma_x$ is given by
\begin{equation}
\label{eq:xsecx}
\sigma_x = f_{12} \frac{\pi e^2}{m_e c \Delta\nu_{\mathrm{D}}} \phi(x).
\end{equation}
Here $f_{12}$ is the Ly$\alpha$ oscillator strength, $e$ and $m_e$ is the
charge and the mass of the electron, and $c$ is the speed of light.
The wavelength dependence of the cross-section enters through the
line profile $\phi(x) = H(a,x)/\sqrt{\pi}$, where $H(a,x)$ is the
Voigt function --- the convolution of a thermal and a natural
broadening of the line --- with $a = \Delta\nu_L / 2\Delta\nu_D$ being the
ratio between the natural and (twice) the thermal line width (the ``damping
parameter'').

In principle, the summation term in Eq.~(\ref{eq:nd}) should also include
a term accounting for the fact that the dust-to-metal ratio $f_{\mathrm{dm}}$
in a given cell may be different from that for which the empirical
data exist. In the Milky Way and the Magellanic Clouds,
$f_{\mathrm{dm}} \simeq 1$
for most metals, i.e.~roughly 1/2 of the metals is condensed to dust grains.
The depletion patterns in high-redshift galaxies are not well
constrained, but no measurements suggest that it should be substantially
different from the local Universe. In fact \citet{pei98} interpret the
depletion patterns of Cr and Zn measured in damped Ly$\alpha$ systems by
\citet{pet97} as giving $f_{\mathrm{dm}} \simeq 1$ for $z \lesssim 3$.
Similarly, fitting depletion patterns of eight elements in GRB host galaxies,
\citet{sav03} find $f_{\mathrm{dm}} \simeq 1$.

\subsubsection{Dust in ionized gas}
\label{sec:iongas}

Ionized gas is found in a number of physically distinct locations throughout
the Universe. Compact
H{\sc ii} regions, or Str\"omgren spheres, surround young, hot stars, while
more diffuse H{\sc ii} is a part of the ISM. Larger H{\sc ii}
``bubbles'' are formed around regions of massive star formation due not only
to ionizing radiation from the stars but also to the energy deposited in the
ISM from supernova feedback. Outside the galaxies, the
IGM is predominantly ionized at $z \lesssim 5$--6.
Observations show or indicate the presence of dust in all of these media.
While
generally lower than in the neutral gas, inferred dust-to-gas mass ratios
($f_{\mathrm{dg}}$) in
ionized gas span a range from roughly equal to the typically assumed MW ISM
value of $\sim$0.01, to upper limits of $\sim$$10^{-4}$ times lower than this.

Based on 45--180 $\mu$m (FIR) spectroscopy, \citet{aan01} found the Galactic
H{\sc ii} region S125 to be strongly depleted of dust, with a dust-to-gas ratio
of $f_{\mathrm{dg}} \le 10^{-6}$, while \citet{smi99}, using MIR imaging and
spectroscopy, inferred a dust-to-gas ratio of the Galactic H{\sc ii} region
RCW 38 of $10^{-5}$ to $10^{-4}$.
On the other hand, using FIR spectroscopy \citet{chi86} found 12 H{\sc ii}
regions to be dust-depleted by ``only'' a factor of 10 relative to the MW ISM
(i.e.~$f_{\mathrm{dg}} \sim 10^{-3}$), while from FIR photometry,
\citet{har71} found the median dust-to-ionized-gas ratio of seven H{\sc ii}
regions to be close to 0.01.

For the more diffuse H{\sc ii} gas that comprises part of the ISM, most
obtained
extinction curves in a sense already include the contribution of H{\sc ii} to
$n_{\mathrm{d}}$, although its quantity is
not revealed when measuring H{\sc i} column densities. Hence, any value of
$f_{\mathrm{ion}}$ for the diffuse ISM
larger than 0 would account twice for the ionized gas. 

The dominant destruction mechanism of dust is probably shock waves, associated
with, e.g., high-velocity clouds and SN winds \citep{dra79a,dra79b}.
However, since SNe are thought to be the prime creator of dust at high
redshifts, the H{\sc ii} bubbles in the vicinity of massive star-forming
regions cannot be entirely void of dust, and observational evidence of dust
related to SN remnants (SNR) and starburst regions does indeed exist.
Using MIR imaging, \citet{bou06}
determined the dust-to-gas ratio of SN 1987A to be $\sim$$5\times10^{-3}$.
Somewhat lower results are found in Kes 75
\citep[$\sim$$10^{-3}$ from FIR and X-ray,][]{mor07} and in Kepler's SN
\citep[$\sim$$10^{-3}$ from IR and bremsstrahlung,][]{con04}.
On larger scales, the hostile environments imposed by the SNe and ionizing
radiation will reduce the dust density in starburst regions. Fitting continuum
SEDs, \citet{con03} found that
$10^{-4} \lesssim f_{\mathrm{dg}} \lesssim 10^{-2}$
in various starburst regions in a sample of seven luminous infrared galaxies.
However, such regions are not ionized to the same level as compact H{\sc ii}
regions and SNRs, and as argued in the case of the diffuse H{\sc ii}, the
scaling of dust with H{\sc i} to some extend already accounts for the
H{\sc ii}.

Various feedback processes are also responsible for expelling a non-vanishing
amount of metals and
dust into the IGM, although inferred dust-to-gas ratios tend to be small:
from IR-to-X-ray luminosites, \citet{gia08} inferred a dust-to-gas ratio of a
few to 5 times $10^{-4}$, as did \citet{che07} by comparing photometric and
spectroscopic properties of quasars behind SDSS clusters.
Higher \citep[dust-to-H{\sc i} $\sim0.05$ in the M81 Group,][]{xil06} ---
possibly expelled from the starburst galaxy M82 --- and lower
\citep[$f_{\mathrm{dg}}\sim10^{-6}$ in the Coma cluster and even less in five
other Abell clusters,][]{sti02}
values are also found. Additionally, sputtering by the hot
halo gas may tend to destroy primarily small grains, leading to a flattening of
the extinction curve in the UV; at the Ly$\alpha$ wavelength, this may reduce
the average cross-section by a factor of 4--5 \citep{agu01}.

In summary, the factor $f_{\mathrm{ion}}$ is a practical way of modeling the
destruction of dust in physically ``hostile'' environments.
For simplicity, in the RT code we will not distinguish between H{\sc ii} in
various regions but merely settle on an average dust-to-gas ratio of ionized
gas of $\sim10^{-4}$; that is we set
$f_{\mathrm{ion}} = 0.01$. In
\S\ref{sec:parstud}, we investigate the effect of other values of
$f_{\mathrm{ion}}$ and find that using 0.01,
the resulting escape fractions lie approximately midway between those
found when using $f_{\mathrm{ion}} = 0$ (corresponding to the \emph{complete}
destruction of dust in regions where hydrogen is ionized) and
$f_{\mathrm{ion}} = 1$ (corresponding to no destruction of dust at all).
Moreover, these extreme values does not seem to change $f_{\mathrm{esc}}$
by more than $\sim25$\%.


\subsection{Albedo}
\label{sec:alb}

When a photon interacts with a dust grain, it may be either absorbed or
scattered. The efficiency with which the dust grain
scatters radiation is dependent on the composition (material, shape, etc.)
of the dust and on the wavelength of the incident photon. If the photon is not
scattered (i.e.~emitted with the same wavelength as the incident photon), it is
absorbed. In this case it is
converted into heat and re-emitted at a later time as infrared radiation.
Expressing the total cross-section as a sum of a scattering cross-section
$\sigma_{\mathrm{s}}$ and an absorbing cross-section $\sigma_{\mathrm{a}}$,
such that $\sigma_{\mathrm{d}} = \sigma_{\mathrm{s}} + \sigma_{\mathrm{a}}$,
the albedo $A$ of the dust is defined as
\begin{equation}
\label{eq:alb}
A = \frac{\sigma_{\mathrm{s}}}{\sigma_{\mathrm{d}}}.
\end{equation}

The albedo of dust has been investigated observationally from reflection
nebulae \citep[e.g.][]{cal95} and diffuse galactic light \citep[e.g.][]{lil76}.
At the Ly$\alpha$ wavelength, $A$ lies approximately between $0.3$ and $0.4$
for various size distributions fitted to the LMC and SMC, assuming that the
dust is made mainly of graphite and silicates \citep{pei92,wei01}.
We adopt an albedo of $A = 0.32$ \citep[from][]{li01}, but investigate the
impact of using other values in \S\ref{sec:parstud}.


\subsection{Phase function}
\label{sec:phase}

If a photon is not absorbed, it is scattered. The probability distribution of
deflection angles $\theta$
from its original path is given by the phase function. For reasons of
symmetry, the scattering must be symmetric in the azimuthal angle $\phi$
(unless the grains are collectively oriented in some preferred direction due
to, e.g., magnetic field lines), but
in general this is not the case in $\theta$. In fact, dust is often
observed to be considerably forward scattering
[e.g.~in reflection nebulae \citep{bur02}, diffuse galactic light
\citep{sch01}, and interstellar clouds \citep{wit90}].
This asymmetric scattering may be described by the \citet{hen41} phase function 
\begin{equation}
\label{eq:Phg}
P_{\mathrm{HG}}(\mu) = \frac{1}{2} \frac{1 - g^2}{(1 + g^2 - 2g\mu)^{3/2}},
\end{equation}
where $\mu = \cos\theta$ and $g = \langle \mu \rangle$ is the asymmetry
parameter. For $g = 0$, Eq.~(\ref{eq:Phg}) reduces to isotropic scattering,
while $g = 1$ $(-1)$ implies complete forward (backward) scattering.
$g$ is a function of
wavelength, but for $\lambda$ close to that of Ly$\alpha$, \citet{li01} found
that $g = 0.73$. Again, other values are investigated in \S\ref{sec:parstud}.



\section{Implementation in Monte Carlo code}
\label{sec:imp}

The numerical code, {\sc MoCaLaTA}, used to conduct the radiative transfer of
Ly$\alpha$, is described in detail in \citet{lau09}. To understand how
the effect of dust is implemented, a brief summary of the concepts of the code
is given below.

The non-dusty version of the code assumes an arbitrary distribution of neutral
hydrogen density
$n_{\textrm{{\scriptsize H}{\tiny \hspace{.1mm}I}}}$, Ly$\alpha$ emissivity
$L$, gas temperature $T$ and velocity field $\mathbf{v}_{\mathrm{bulk}}$.
The physical parameters, typically resulting from a cosmological simulation,
are assigned in a cell-based structure, where any cell may be refined, i.e.
split up in eight sub-cells, recursively to an arbitrary level of refinement,
thus allowing for investigation of a detailed structure.

The basic principles of the code are as follows: photons are
emitted from a random point in space, weighted according to the emissivity
at that particular point,
i.e.~a given photon has a probability $L_i / L_{\mathrm{tot}}$ of being emitted
from the $i$'th cell, where $L_{\mathrm{tot}}$ is the total Ly$\alpha$
luminosity of the system. It is emitted in a random direction, with a frequency
close to the Ly$\alpha$ line center.

The photon travels through the gas, traversing an optical depth $\tau$ with a
probability $e^{-\tau}$, before it is scattered on a hydrogen atom.
Re-emitted in a new,
random direction, weighted according to an appropriate phase function, it
continues its journey until it escapes the computational box. At each
scattering, the velocity (thermal + bulk) of the
atom will Doppler shift the frequency of the
photon, so that it diffuses not only in real but also in frequency space.
This procedure
is then repeated for a number $n_{\mathrm{ph}}$ of photons sufficient to
provide good statistic for the quantities of interest (several 1000s, or
millions, depending on the desired output).

For each photon and for each scattering, the probability that the photon
escape in the directions of six virtual observers located at a distance of the
luminosity distance of the galaxy along the three Cartesian axes is computed
and added to an associated array (a ``CCD'') of three dimensions; two spatial
and one frequential, so that a full spectrum is obtained for each pixel in the
image.

\subsection{Metallicity normalization}
\label{sec:znorm}

In addition to the parameters
$n_{\textrm{{\scriptsize H}{\tiny \hspace{.1mm}I}}}$, $L$, $T$ and
$\mathbf{v}_{\mathrm{bulk}}$, every cell has an ionized hydrogen density
$n_{\textrm{{\scriptsize H}{\tiny \hspace{.1mm}II}}}$ and a metallicity $Z_i$
of the different elements associated with it, from which a dust density (per
hydrogen atom) $n_{\mathrm{d}}$ is calculated according to Eq.~(\ref{eq:nd}).


With a randomly drawn optical depth to be covered by a photon, the distance $r$
traveled can then be calculated as
\begin{equation}
\label{eq:r}
r = \frac{\tau}{n_{\textrm{{\scriptsize H}{\tiny \hspace{.1mm}I}}} \sigma_x
  +             n_{\mathrm{d}} \sigma_{\mathrm{d}}}.
\end{equation}

Having reached this distance, a random number $\mathcal{R} \in [0,1]$
(a ``univariate'') determines
whether the photon hits a hydrogen atom or a dust grain by comparing it to the
ratio $\varrho = n_{\mathrm{d}} \sigma_{\mathrm{d}} /
(n_{\textrm{{\scriptsize H}{\tiny \hspace{.1mm}I}}} \sigma_x +
n_{\mathrm{d}} \sigma_{\mathrm{d}}) =
\tau_{\mathrm{d}} / (\tau_x + \tau_{\mathrm{d}})$;
if $\mathcal{R} \le \varrho$, the interaction
is caused by dust. In this case a second univariate is compared to the
albedo of the dust grain, dictating whether the the photon is absorbed, thus
terminating the journey of this particular photon, or scattered, in which case
it is re-emitted in a random direction given by Eq.~(\ref{eq:Phg}).

Various acceleration schemes were implemented in the non-dusty Ly$\alpha$ RT;
the extension of these to the dusty version is discussed in App.~\ref{sec:acc}.



\section{Testing the code}
\label{sec:test}

\begin{figure}
\epsscale{1.}
\plotone{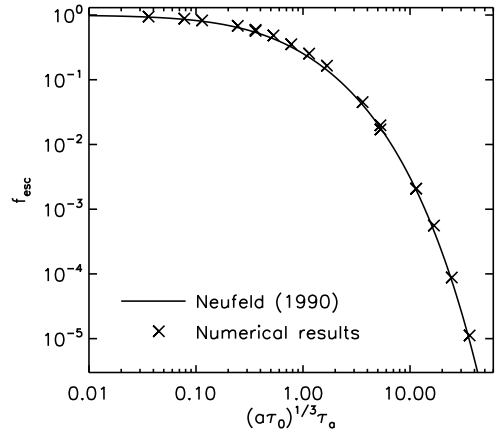}
\caption{Escape fractions $f_{\mathrm{esc}}$ of photons emitted from the center
         of a semi-infinite slab of gas damping parameter $a$, hydrogen optical
         depth $\tau_0$, and dust absorbing optical depth $\tau_{\mathrm{a}}$,
         compared to the analytical solution in Eq.~(\ref{eq:neufesc}).}
\label{fig:neufesc}
\end{figure}

\citet{neu90} provided an analytical expression for the escape fraction of
photons emitted from inside a ``slab'' (i.e.~finite in the $z$-direction and
infinite in the $xy$-direction) of an absorbing medium. The solution, which is
valid for very high optical depths ($a\tau_0 \gtrsim 10^3$, where $\tau_0$ is
the optical depth of neutral hydrogen from the center to the surface of the
slab) and in the limit $(a\tau_0)^{1/3} \gg \tau_{\mathrm{a}}$, where
$\tau_{\mathrm{a}}$ is the absorption optical depth of dust, is
\begin{equation}
\label{eq:neufesc}
f_{\mathrm{esc}} = \frac{1}
   {\cosh\left[ \zeta'
                \sqrt{(a\tau_0)^{1/3}\tau_{\mathrm{a}}}\right]},
\end{equation}
where $\zeta' \equiv \sqrt{3} / \zeta\pi^{5/12}$,
with $\zeta \simeq 0.525$ a fitting parameter. Figure \ref{fig:neufesc}
shows the result of a series of such simulations, compared to the analytical
solution.

Various AMR grid configurations were tested. Of course, the physical parameters
such as $n_{\textrm{{\scriptsize H}{\tiny \hspace{.1mm}I}}}$, $n_{\mathrm{d}}$,
and $T$ of a cell does not depend on the refinement level, but the acceleration
schemes discussed in App.~\ref{sec:acc} do.


\section{Simulations}
\label{sec:sim}

To investigate the fraction of emitted Ly$\alpha$ photons that escapes galaxies
at high redshift, the code was applied to a number of different simulated
high-resolution galaxies, emerging from a fully cosmological
N-body/hydrodynamical TreeSPH simulation. This code was described in detail in
\citet{som03} and \citet{som06}, and summarized in \citet{lau09}. Here, we will
only briefly review the basic characteristics of the code:

The numerical simulations are first carried out at low resolution, but in a
large volume of space. Subsequently, interesting galaxy-forming regions are
re-simulated at high resolution. In addition to hydrogen and helium, the code
follows the chemical evolution of C, N, O, Mg, Si, S, Ca, and Fe.

The Ly$\alpha$ emission is produced by three different processes
\citep[see also][]{lau07},
viz.~from recombinations in photoionized regions around massive stars
(responsible for $\sim$90\% of the total Ly$\alpha$ luminosity),
gravitational cooling of infalling gas ($\sim$10\%), and a metagalactic UV
background (UVB) photoionizing the external parts of the galaxy ($\sim$1\%).

The UVB field is assumed to be that given by \citet{haa96}, where the
gas is treated as optically thin to the UV radiation until the mean free path
of a UV photon at the Lyman limit becomes less than 0.1 kpc, at which point the
gas is treated as optically thick and the UV field is ``switched off''.

A far more realistic UV RT scheme was implemented in the code by
\citet{raz06,raz07} which alters the ionization state and temperature field of
the gas somewhat. However, as was concluded in \citet{lau09}, the effect of the
improved UV RT only results in minor changes for the non-dusty Ly$\alpha$ RT.
Moreover, as shown
in \S\ref{sec:parstud} it does not alter $f_{\mathrm{esc}}$ drastically,
and has thus not been applied in the present paper.

Nine individual galaxies are extracted from
the cosmological simulation at redshift $z = 3.6$ --- at which time the
Universe was 1.8 Gyr old --- to be used for the
Ly$\alpha$ RT. These galaxies
are representative of typical galaxies in the sense that they
span three orders of magnitudes in mass, the most massive
eventually evolving into a disk galaxy with circular speed
$V_{\mathrm{c}} \simeq 300$ km s$^{-1}$ at $z = 0$.
The numerical and physical properties of these galaxies are
listed in Tab.~\ref{tab:num} and Tab.~\ref{tab:phy}, respectively.
\ifnum\astroph=0
  \begin{deluxetable}{lccccccccc}
\else
  \begin{deluxetable*}{lccccccccc}
\fi
\tablecolumns{10}
\tablewidth{0pc}
\tablecaption{Characteristic quantities of the simulations}
\tablehead{
\colhead{Galaxy}                                     &          S33sc    &            K15       &           S29     &            K33        &         S115        &          S87   &          S108  &         S115sc  &          S108sc \\
}
\startdata
$N_{\mathrm{p,tot}}$                                 & 1.2$\times$$10^6$ &   2.2$\times$$10^6$  & 1.1$\times$$10^6$ &  $1.2$$\times$$10^6$  & $1.3$$\times$$10^6$ & 1.4$\times$$10^6$ & 1.3$\times$$10^6$ & 1.3$\times$$10^6$  & 1.3$\times$$10^6$  \\
$N_{\mathrm{SPH}}$                                   & 5.5$\times$$10^5$ &   1.0$\times$$10^6$  & 5.1$\times$$10^5$ &  $5.5$$\times$$10^5$  & $6.4$$\times$$10^5$ & 7.0$\times$$10^5$ & 6.3$\times$$10^5$ & 6.4$\times$$10^5$  & 6.3$\times$$10^5$  \\
$m_{\mathrm{SPH}}$,$m_{\mathrm{star}}$               & 5.4$\times$$10^5$ &   9.3$\times$$10^4$  & 9.3$\times$$10^4$ &  $9.3$$\times$$10^4$  & $1.1$$\times$$10^4$ & 1.2$\times$$10^4$ & 1.2$\times$$10^4$ & 2.6$\times$$10^3$  & 1.5$\times$$10^3$  \\
$m_{\mathrm{DM}}$                                    & 3.0$\times$$10^6$ &   5.2$\times$$10^5$  & 5.2$\times$$10^5$ &  $5.2$$\times$$10^5$  & $6.6$$\times$$10^4$ & 6.5$\times$$10^4$ & 6.5$\times$$10^4$ & 1.4$\times$$10^4$  & 8.1$\times$$10^3$  \\
$\epsilon_{\mathrm{SPH}}$,$\epsilon_{\mathrm{star}}$ &     344           &         191          &     191           &       191             &      96             &    96             &        96         &     58             &     48             \\
$\epsilon_{\mathrm{DM}}$                             &     612           &         340          &     340           &       340             &      170            &   170             &       170         &    102             &     85             \\
$l_{\mathrm{min}}$                                   &      18           &         10           &      10           &        10             &        5            &     5             &         5         &      3             &      2.5           \\
\enddata
\tablecomments{Total number of particles ($N_{\mathrm{p,tot}}$), number
               of SPH particles only ($N_{\mathrm{SPH}}$), masses ($m$),
               gravity
               softening lengths ($\epsilon$), and minimum smoothing lengths
               ($l_{\mathrm{min}}$) of dark matter (DM), gas (SPH), and star
               particles used in the simulations. Masses are measured in
               $h^{-1}M_\odot$, distances in $h^{-1}$pc.}
\label{tab:num}
\ifnum\astroph=0
  \end{deluxetable}
\else
  \end{deluxetable*}
\fi
\ifnum\astroph=0
  \begin{deluxetable}{lccccccccc}
\else
  \begin{deluxetable*}{lccccccccc}
\fi
\tablecolumns{10}
\tablewidth{0pc}
\tablecaption{Physical properties of the simulated galaxies}
\tablehead{
\colhead{Galaxy}                             &           S33sc        &            K15         &           S29        &            K33         &          S115          &          S87           &          S108         &          S115sc         &          S108sc      \\
}                                                                                                                                                                                                                                                                         
\startdata                                                                                                                                                                                                                                                                
SFR/$M_\odot$ yr$^{-1}$                      &         70             & 16                     &           13         & 13                     & 0.5                    &        0.46            &        1.62           &  $3.7$$\times$$10^{-3}$ & $1.7$$\times$$10^{-3}$ \\
$M_*/M_\odot$                                & $3.4$$\times$$10^{10}$ & $1.3$$\times$$10^{10}$ & $6.0\times10^{9}$    & $6.5$$\times$$10^{9}$  & $2.5$$\times$$10^{8}$  & $1.8$$\times$$10^{8}$  & $4.9$$\times$$10^{8}$ &  $2.0$$\times$$10^{7}$  & $5.9$$\times$$10^{6}$  \\
$M_{\mathrm{vir}}/M_\odot$                   & $7.6$$\times$$10^{11}$ & $2.8$$\times$$10^{11}$ & $1.7\times10^{11}$   & $1.3$$\times$$10^{11}$ & $2.5$$\times$$10^{10}$ & $2.1$$\times$$10^{10}$ & $2.6$$\times$$10^{9}$ &  $4.9$$\times$$10^{9}$  & $3.3$$\times$$10^{8}$  \\
$r_{\mathrm{vir}}$/kpc                       &      63                & 45                     &       39             & 35                     & 20                     &   19                   &     10                &        12               &          5             \\
$[$O/H$]$                                    &    $-0.08$             &     $-0.30$            &   $-0.28$            &     $-0.40$            &    $-1.22$             &    $-1.28$             &   $-0.51$             &     $-1.54$             &       $-1.64$          \\
$V_{\mathrm{c}}(z=0)$/km s$^{-1}$            &   300                  & 245                    &    205               & 180                    & 125                    &           132          &       131             &        50               &            35          \\
$L_{\mathrm{Ly}\alpha}$/erg s$^{-1}$         & 1.6$\times$$10^{44}$   & $4.5$$\times$$10^{43}$ & 2.9$\times$$10^{43}$ & $2.5$$\times$$10^{43}$ & $1.3$$\times$$10^{42}$ & 1.1$\times$$10^{42}$   & 2.6$\times$$10^{42}$  & 4.9$\times$$10^{40}$    & 3.2$\times$$10^{39}$   \\
$L_{\nu,\mathrm{UV}}$/erg s$^{-1}$ Hz$^{-1}$ & 5.0$\times$$10^{29}$   & $6.7$$\times$$10^{28}$ & 9.3$\times$$10^{28}$ & $5.5$$\times$$10^{28}$ & $3.6$$\times$$10^{27}$ & 3.3$\times$$10^{27}$   & 1.2$\times$$10^{28}$  & 2.6$\times$$10^{25}$    & 1.2$\times$$10^{25}$   \\
\enddata
\tablecomments{Star formation rates (SFRs),
               stellar masses ($M_*$),
               virial masses ($M_{\mathrm{vir}}$),
               virial radii ($r_{\mathrm{vir}}$),
               metallicities ($[$O/H$]$),
               circular velocites ($V_{\mathrm{c}}$),
               Ly$\alpha$ luminosities ($L_{\mathrm{Ly}\alpha}$),
               and UV luminosities ($L_{\nu,\mathrm{UV}}$)
               for the simulated galaxies.
               All quoted values correspond to a redshift of
               $z = 3.6$, except $V_{\mathrm{c}}$ which is given for $z = 0$.}
\label{tab:phy}
\ifnum\astroph=0
  \end{deluxetable}
\else
  \end{deluxetable*}
\fi

Since the Ly$\alpha$ RT code is
cell-based, the relevant physical properties of the particles are interpolated
from the 50 nearest neighboring particles onto a grid of base resolution
$128^3$ cells.
Dense cells are subdivided in eight cells, which are further
refined until no cell contains more than ten particles. Achieving the same
resolution without this adaptively refined mesh would require from
$\sim$$16\,000^3$ to $\sim$$65\,000^3$ cells; infeasible with
present-day's computer facilities.

\begin{figure*}
\epsscale{1}
\plotone{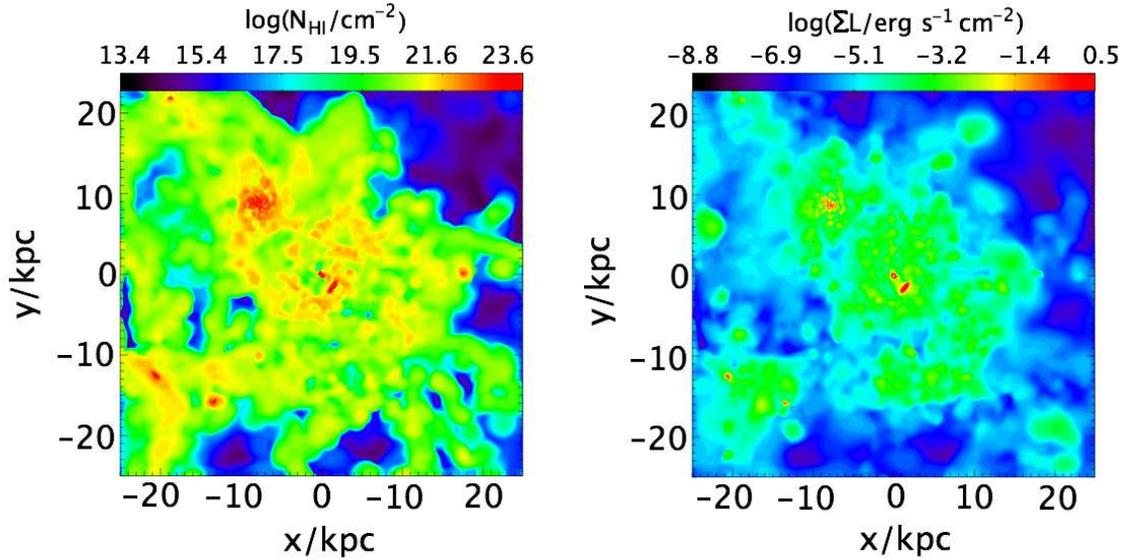}
\caption{Neutral hydrogen column density
         ($N_{\textrm{{\scriptsize H}{\tiny \hspace{.1mm}I}}}$)
         map (\emph{left}) and integrated source Ly$\alpha$ emissivity 
         ($\Sigma L$) map (\emph{right}) of the simulated galaxy K15.
         The vast majority of the photons are seen to be emitted in the very
         dense environments. According to the analytical solution for the
         Ly$\alpha$ escape fraction provided by
         \citet[][Eq.~(\ref{eq:neufesc}) in this paper]{neu90}, virtually all
         photons should be absorbed by dust, but taking into account the
         clumpiness of the ISM allows for much higher escape fractions.}
\label{fig:NHI_L}
\end{figure*}

This level of resolution is rather crucial to the escape of the Ly$\alpha$
photons. As is evident from Fig.~\ref{fig:NHI_L},
the bulk of the photons is produced in regions of very high column densites,
exceeding
$N_{\textrm{{\scriptsize H}{\tiny \hspace{.1mm}I}}} = 10^{23}$ cm$^{-2}$.
Even for metallicities of only $1/100$ $Z_\odot$, Eq.~(\ref{eq:neufesc})
implies an escape fraction of $\sim$$10^{-5}$ (for $T \sim 10^4$ K). However,
Eq.~(\ref{eq:neufesc}) assumes a \emph{homogeneous} medium. As was argued by
\citet{neu91} and investigated numerically by \citet{han06}, in a multi-phase
medium, the Ly$\alpha$ photons may escape more easily. In the (academic) case
of all the dust residing in cool, dense clouds of neutral hydrogen which, in
turn, are dispersed in a hot, empty medium, the Ly$\alpha$ escape
fraction may approach unity. The reason is that the photons will scatter off of
the surface before penetrating substantially into the clouds, thus effectively
confining their journey to the dustless intercloud medium.

Although this scenario is obviously very idealized, the presence of an
inhomogeneous medium undoubtedly reduces the effective optical depth, and
has indeed been invoked to explain unusually large Ly$\alpha$ equivalent widths
\citep[e.g.][]{fin08}. Nevertheless, it is not clear to which degree the
escape fraction of Ly$\alpha$ will actually be affected.
Other scenarios have been proposed to explain the apparent paradoxal escape of
Ly$\alpha$, e.g.~galactic superwinds, as discussed in the introduction.
Here we present the first calculations of $f_{\mathrm{esc}}$, based on fully
cosmological, numerical galaxy formation models, demonstrating that generally
of the order 5\%--30\% of the Ly$\alpha$ radiation escapes the galaxies at
redshifts $z \sim 3$--4, even if no particularly strong winds are present.


\section{Results}
\label{sec:res}

In the discussion of the results, we will first focus on one particular galaxy,
observed from one direction. We arbitrarily chose K15, which at $z = 0$ becomes
an M31-sized disk galaxy, and we chose the observer placed
toward negative values on the $z$-axis (the $z_-$-direction),
the direction in which most radiation escapes. In \S\ref{sec:moregal} we
proceed to discuss the extend to which the results are representative for other
galaxies and directions.

\subsection{Where are the photons absorbed?}
\label{sec:where}

\begin{figure*}
\epsscale{1}
\plotone{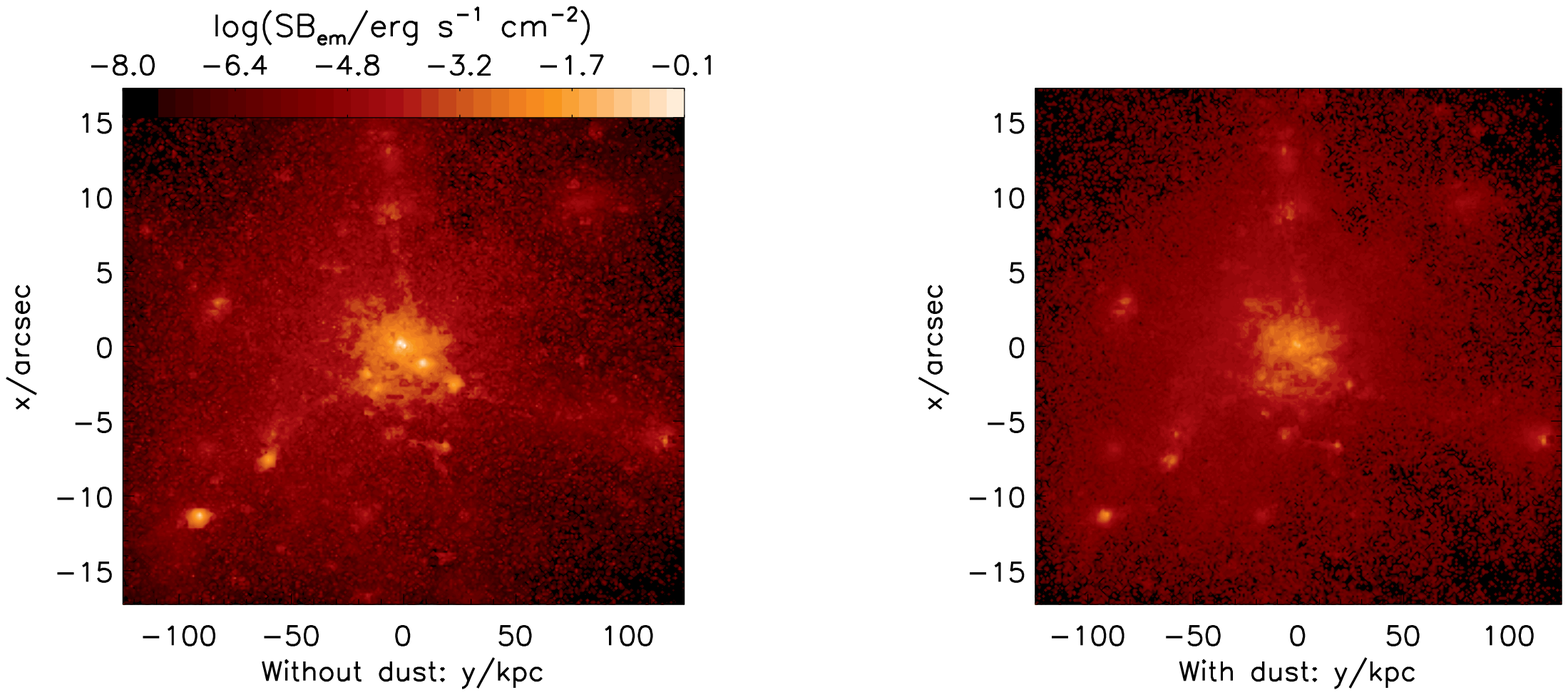}
\caption{Surface brightness maps of the galaxy K15, as viewed from the
         negative $z$-direction, without dust (\emph{left}) and with dust
         (\emph{right}). Including dust in the radiative transfer affects
         primarily the most luminous regions.}
\label{fig:SBmap}
\end{figure*}

\begin{figure}
\epsscale{1}
\plotone{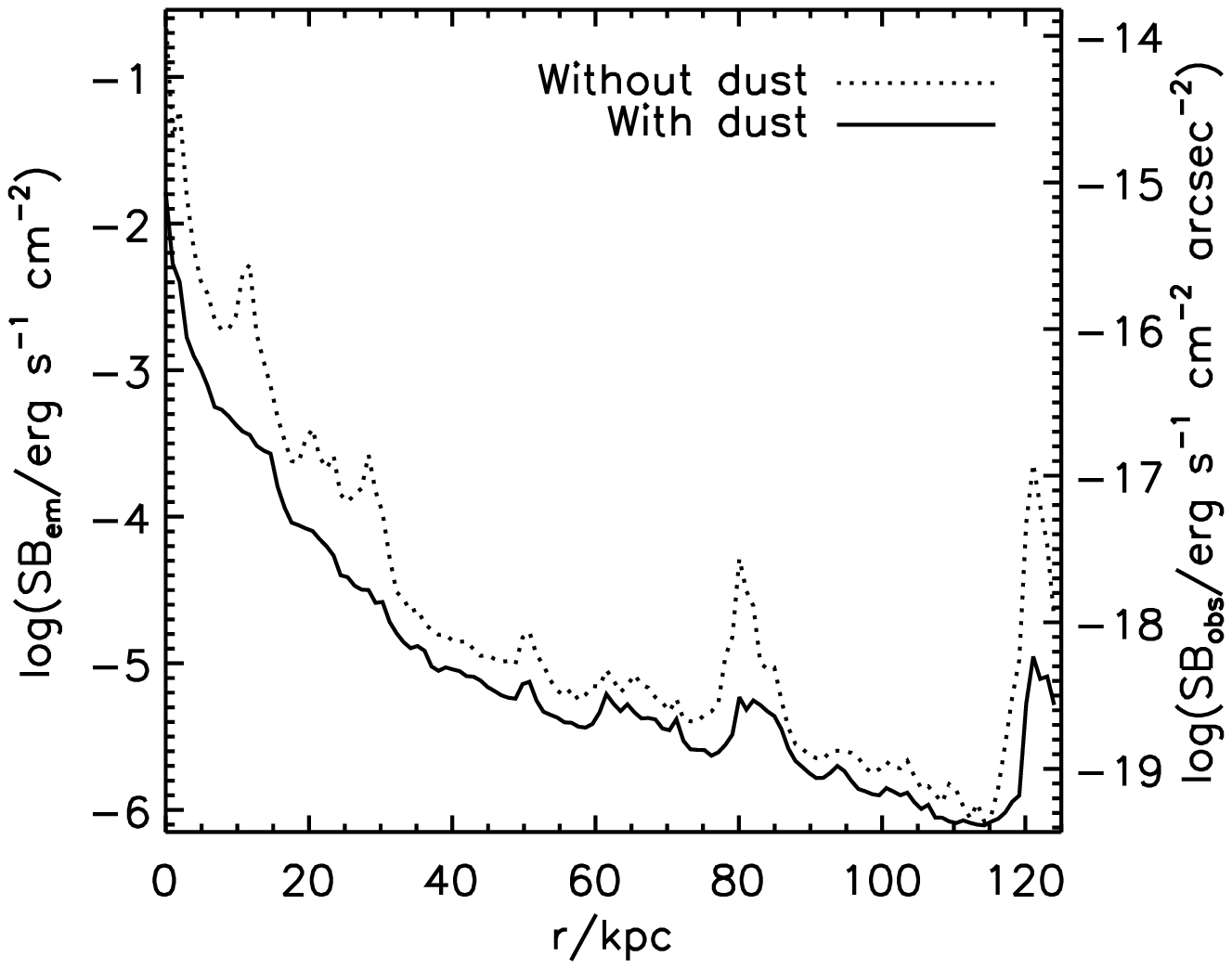}
\caption{Surface brightness (SB) profile of the galaxy K15, again with dust
         (\emph{solid}) and without dust (\emph{dotted}).
         Left ordinate axis gives the SB as measured at the source while right
         ordinate gives the values measured by an observer at a distance given
         by the luminosity distance of the galaxy.
         The decrease in SB in the less luminous regions
         is noticable. However, this decrease is for the most part \emph{not}
         due to photons being
         absorbed in the hot and tenuous circumgalactic medium but rather
         reflects a lack of photons that in the case of no dust would have
         escaped the luminous regions and subsequently scattered on neutral
         hydrogen in the direction of the observer.}
\label{fig:SBprof}
\end{figure}

Figure \ref{fig:SBmap} shows the surface brightness (SB) maps of the galaxy
K15. The left panel shows how the galaxy would look if the gas were dust-free,
while the right panel displays the more realistic case of a dusty medium.
Comparing the two images, it is seen that the regions that are
affected the most by dust are the most luminous regions. This is even more
evident in Fig.~\ref{fig:SBprof}, where the azimuthally averaged profiles of
the SB maps are shown.

The reason for this is two-fold: the most luminous regions are the regions
where the stars reside. Because dust is produced by stars, this is also where
most of the dust is. Since the stars, in turn, are born in regions of high
hydrogen density, the RT is here associated with numerous
scatterings, severely increasing the path length of the photons, and
consequently increasing further the probability of being absorbed.

\begin{figure}
\epsscale{1}
\plotone{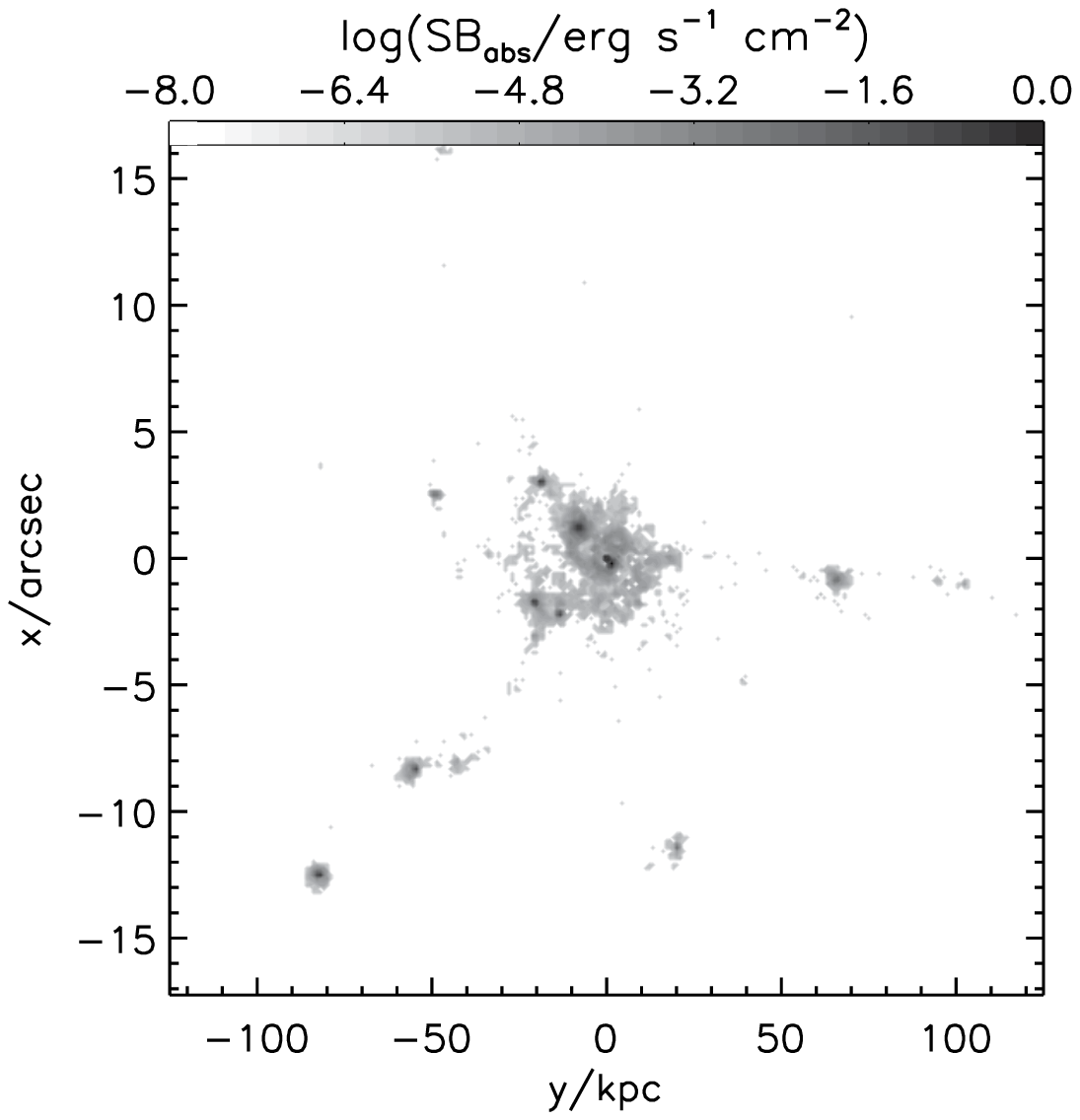}
\caption{Image of the locations of absorption of the Ly$\alpha$ radiation that
         does \emph{not} escape K15.
         Effectively, this image shows the column density of
         dust.}
\label{fig:IR}
\end{figure}

The SB in the less luminous regions also decreases. Although dust also resides
here, having been expelled by the feedback of starbursts, only little
absorption actually takes place here. The decrease in luminosity is mainly
due to the photons being absorbed in the high density regions that would
otherwise have escaped the inner regions and subsequently been scattered by the
circumambient neutral gas into the direction of the observer. This is also
discernible from Fig.~\ref{fig:IR} that displays an image of the absorbed
photons.

\begin{figure*}
\epsscale{1}
\plottwo{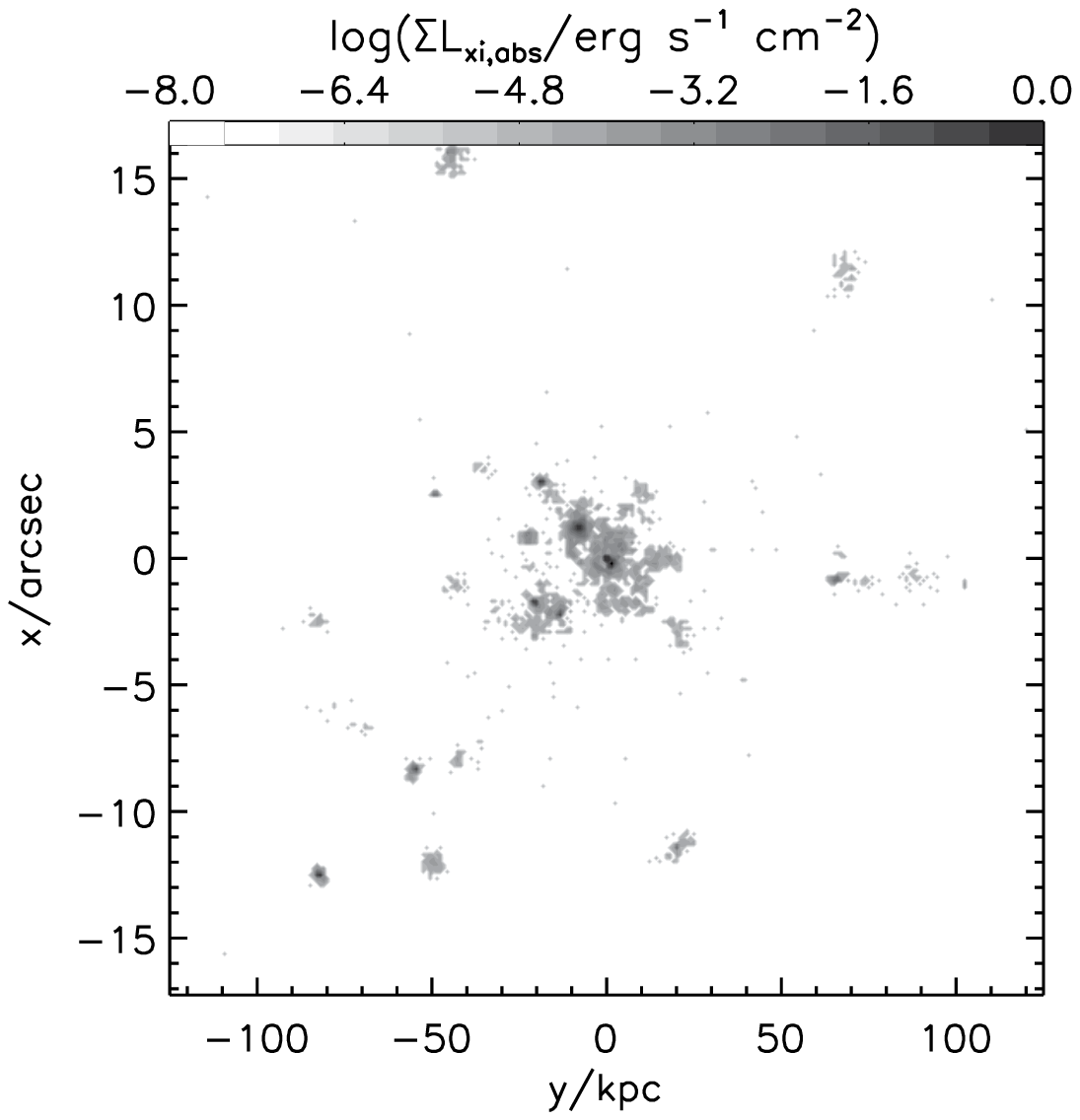}{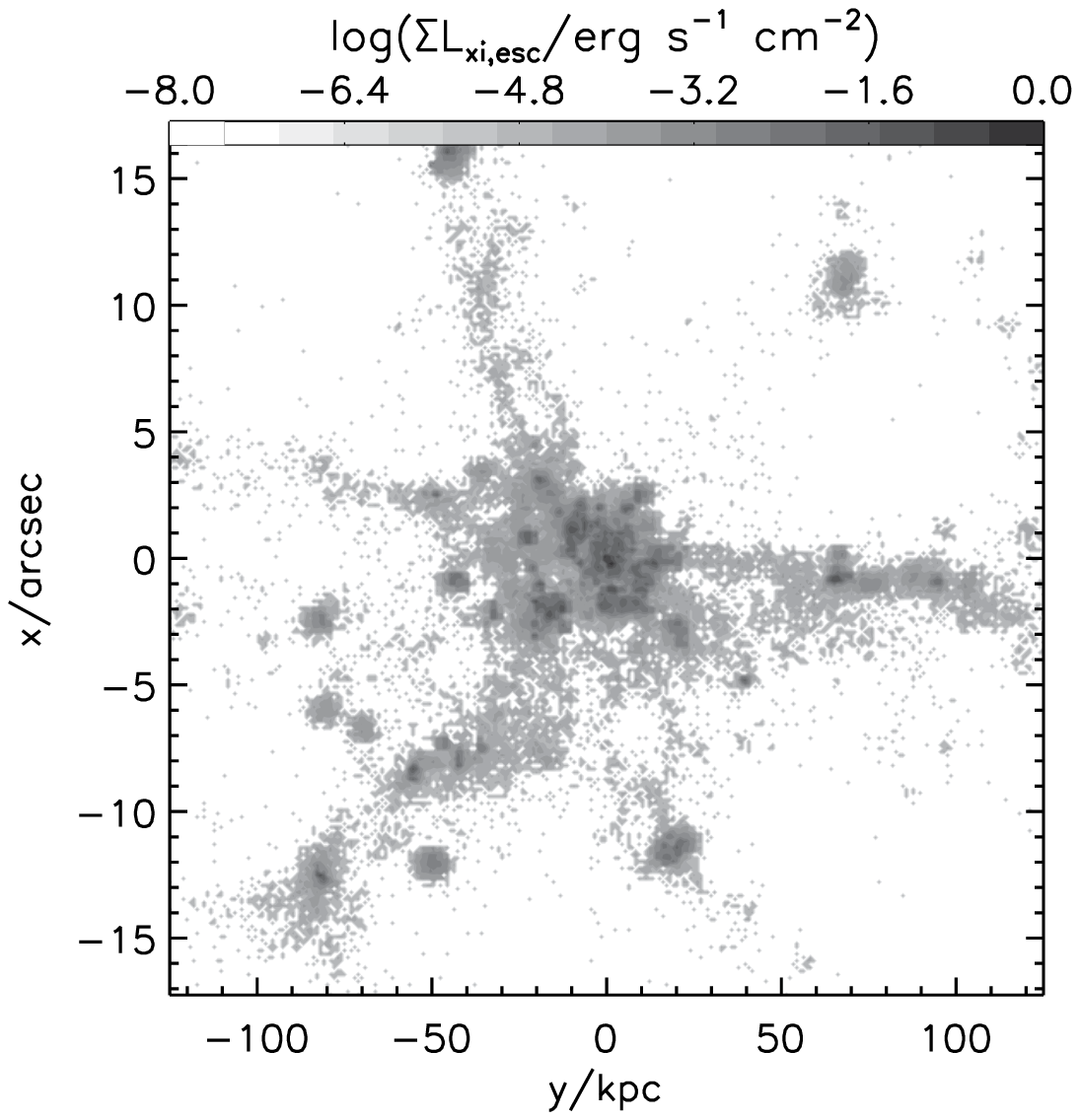}
\caption{Surface brightness maps of the source Ly$\alpha$ emissivity, of
         the photons that are eventually absorbed (\emph{left}), and the photons
         that eventually escape (\emph{right}). These images clearly show that
         absorbed and escaping photons do not in general probe the same
         physical domains.}
\label{fig:Xi}
\end{figure*}

Figure \ref{fig:Xi} shows the \emph{source} Ly$\alpha$ emissivity of the
photons that are eventually absorbed, compared to that of the photons that
eventually escape. Here it becomes evident that virtually all the absorbed
photons are emitted from the central parts, while photons that escape are
emitted from everywhere. In particular, the radiation produced through
gravitational cooling escapes more or less freely.

%


\subsection{Effect on the emergent spectrum}
\label{sec:spec}

\begin{figure}
\epsscale{1}
\plotone{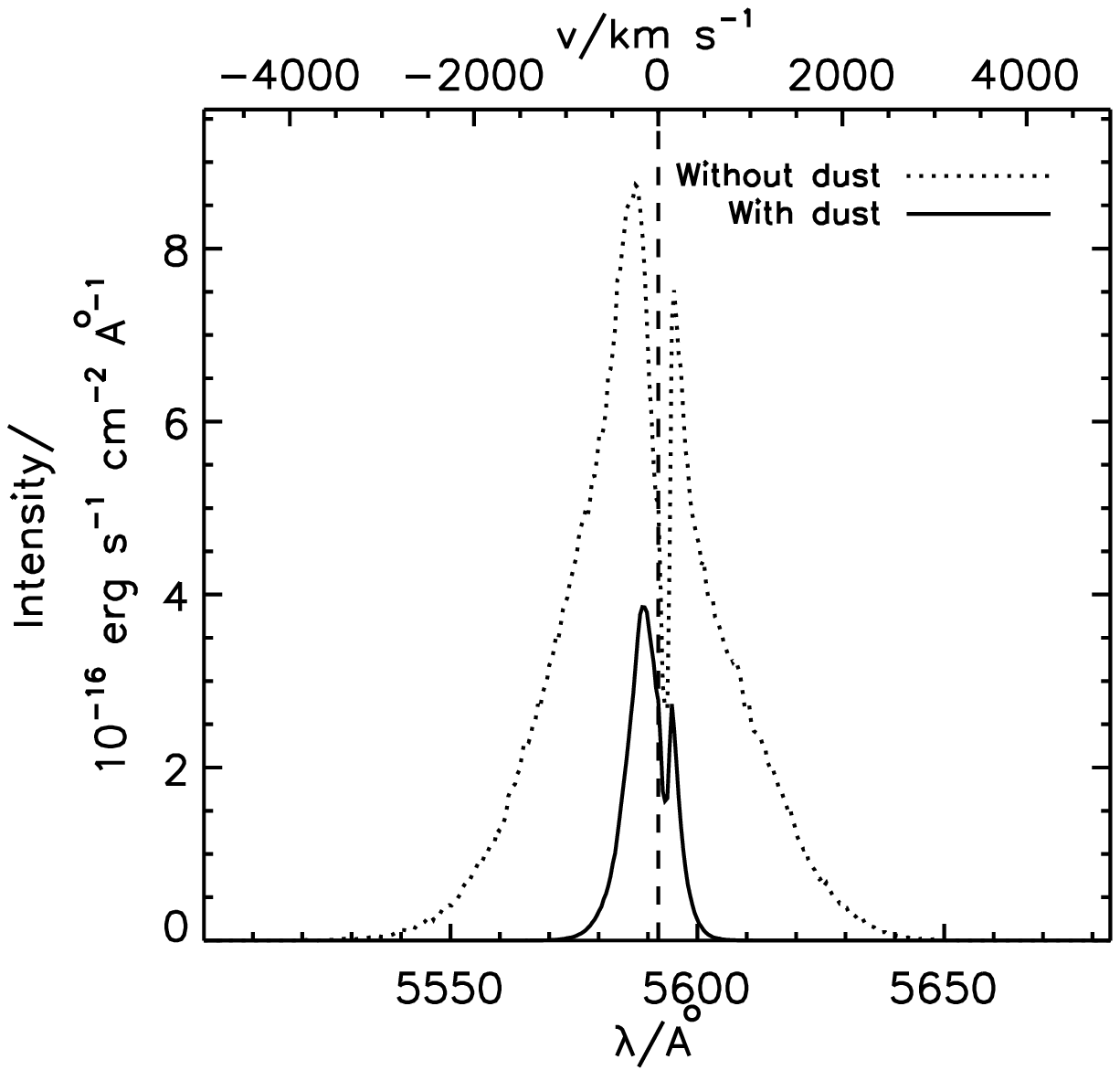}
\caption{Observed spectral distribution of the radiation escaping the galaxy
         K15 in the negative $z$-direction, with (\emph{solid}) and without
         (\emph{dotted}) dust. The vertical, dashed line marks the Ly$\alpha$
         line center. Although the dust is close to being grey, the
         spectrum is not affected in the same way at all wavelengths; the inner
         parts are only diminished by a factor of $\sim$2, while the wings are
         severely reduced. The reason is that the photons in the wings are the
         ones produced in the dense and dusty regions of the galaxy, so these
         photons have a higher probability of being absorbed.}
\label{fig:spec}
\end{figure}

Due to the high opacity of the gas for a Ly$\alpha$ photon at line center,
photons generally  diffuse in frequency to either the red or the blue side in
order to facilitate escape.
Consequently, the spectrum of the radiation escaping a
dustless medium is characterized by a double-peaked profile. The broadening of
the wings is dictated by the product $a\tau_0$ \citep{har73}, i.e.~low
temperatures and, in particular, high densities force the photons to diffuse
far from line center. Since such conditions are typical of the regions where
the bulk of the photons is absorbed, the emergent spectrum of a dusty medium
is severely narrowed, although the double-peaked feature persists.
Figure \ref{fig:spec} displays the spatially integrated spectra of the dustless
and the dusty version of K15.

This interesting result shows that even though the dust is effectively grey
(the small wavelength dependence of Eq.~(\ref{eq:sigd}) does not produce
substantially different results from using a completely wavelength
\emph{in}dependent
cross-section), the emergent spectrum is affected in a highly ``non-grey''
fashion:
whereas the escape fraction of the inner part of the spectrum is of the order
50\%, it rapidly drops when moving away from the line center.


\subsection{Escape fraction}
\label{sec:fesc}

In the $z_-$-direction of K15, the fraction of Ly$\alpha$ photons
escaping is 0.14. As mentioned earlier, the $z_-$-direction is the direction
into which most
radiation escapes --- without dust, this direction is $\sim3$ times as
luminous as the $x_-$-direction, which is where the least radiation is emitted.
Including dust, because the brightest regions are affected the most, the ratio
between the luminosity in the least and the most luminous directions is
somewhat reduced, although $z_-$ is still more than twice as bright as $x_-$.

The sky-averaged escape fraction for K15 is 0.16.
Recall that in these simulations SMC dust
has been applied. However, as discussed in \S\ref{sec:parstud} using LMC dust
does not alter the results significantly.


\subsection{General results}
\label{sec:moregal}

The results found in the previous sections turn out to be quite illustrative
of the general outcome of Ly$\alpha$ RT in a dusty medium: photons are absorbed
primarily in the dense, luminous regions, leading to a reduced luminosity in
these parts of the galaxies and effectively smoothing out prominent features
Furthermore, the wings of the spectrum experience a strong cut-off.

\subsubsection{Anisotropic escape of Ly$\alpha$}
\label{sec:anis}

In general, the radiation does \emph{not} escape isotropically; the ratio of
luminosities observed from different directions ranges from $\sim$1.5 to
$\sim$4. Without dust, these ratios are somewhat higher, up to a factor of
$\sim$6.
Although ``bright directions'' are affected more by the dust than less bright
directions, the variation in $f_{\mathrm{esc}}$ as a function of direction is
not large, and not very different from the sky-averaged $f_{\mathrm{esc}}$.

\subsubsection{Correlation of $f_{\mathrm{esc}}$ with galactic mass}
\label{sec:fesc_M}

\begin{figure}
\epsscale{1}
\plotone{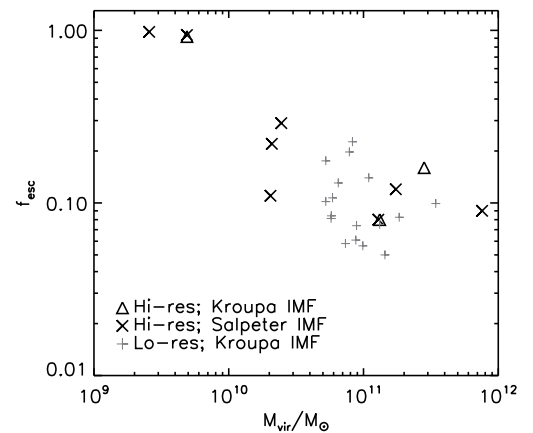}
\caption{Escape fraction $f_{\mathrm{esc}}$ as a function of galactic virial
         mass $M_{\mathrm{vir}}$. Although the plot exhibits a large scatter,
         there is a clear trend of $f_{\mathrm{esc}}$ decreasing with
         increasing $M_{\mathrm{vir}}$.}
\label{fig:fesc_M}
\end{figure}

Figure \ref{fig:fesc_M} shows the escape fractions of the galaxies as a
function of the virial masses of the galaxies. Despite a large scatter, and
although the sample is too small to say anything definite, the figure indicates
that $f_{\mathrm{esc}}$ decreases with increasing mass of the host galaxy.
In addition to the nine high-resolution galaxies [of which two appear in
two ``versions''; with a \citet{sal55} and with a \citet{kro98} initial mass
function (IMF)], 17 galaxies from a
low-resolution (Kroupa) simulation are shown. These galaxies were chosen
according to the criterion that the number of star particles is $\ge$1000 to
ensure an acceptable resolution.
Similar results are found for the escape of ionizing UV radiation
\citep{raz09}. The reason is probably a combination of two mechanisms: small
galaxies have a lower metallicity and hence less dust than large galaxies.
Furthermore, due to their smaller gravitational potential the stellar feedback
will ``puff up'' small galaxies relatively more and make them less ordered,
thus reducing the column density of both dust and gas.

%

\subsubsection{Narrowing of the spectrum}
\label{sec:narrow}

\begin{figure*}
\epsscale{1}
\plotone{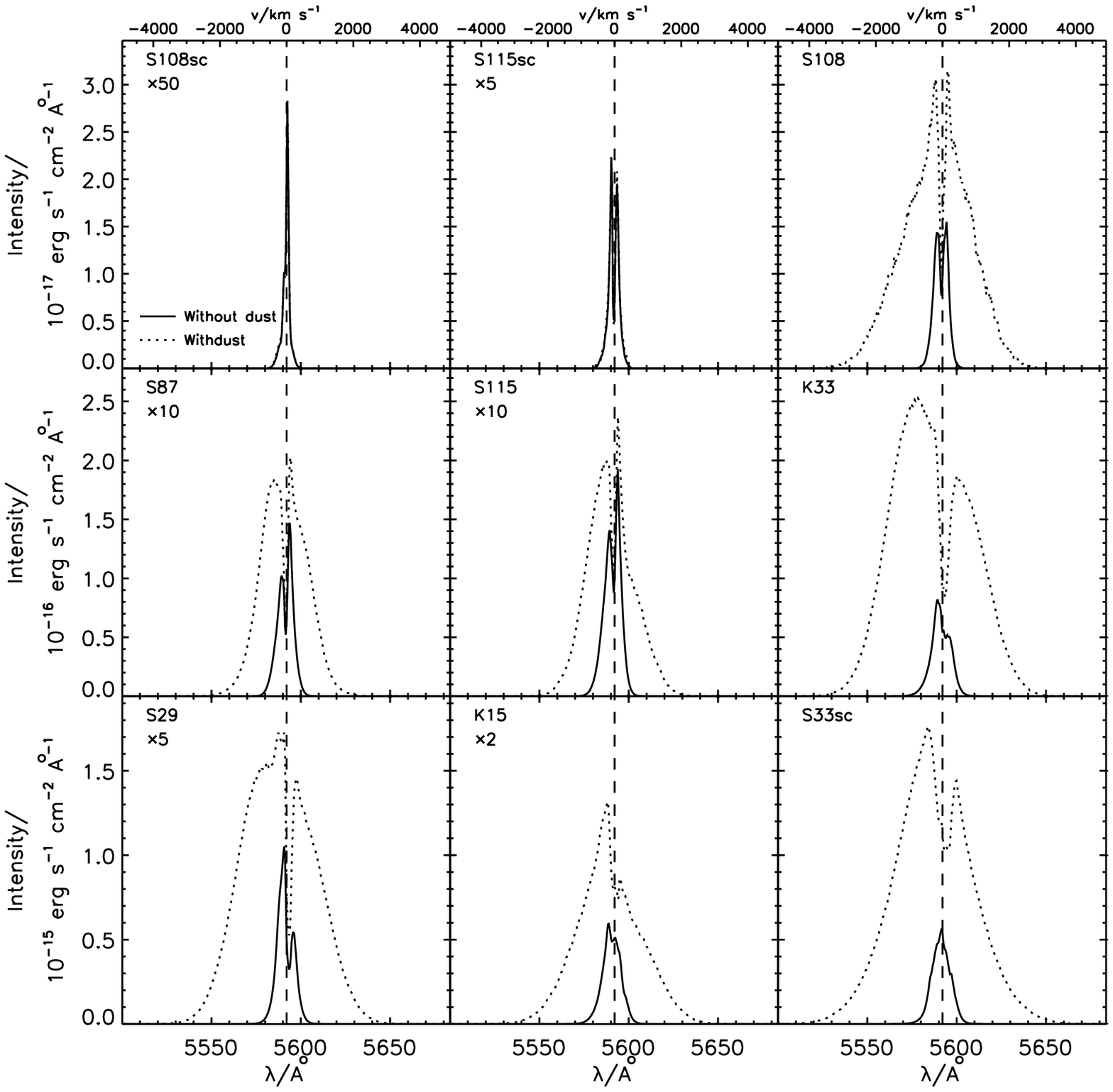}
\caption{Emergent spectra of the studied galaxies, without dust
         (\emph{dotted}) and with dust (\emph{solid}), ordered after decreasing
         virial mass.
         In order to use the same ordinate axis for a given row, some
         intensities have been multiplied a factor indicated under the name of
         the galaxy.
         Generally, lines that are broadened by resonant scattering tend to be
         severely narrowed when including dust.}
\label{fig:SpecPanel}
\end{figure*}

Figure \ref{fig:SpecPanel} shows the emergent spectra of the studied galaxies,
arranged according to virial mass. In general, the same trend as for K15 is
seen for all galaxies: the dust primarily affects the wings of the profiles.
As discussed in \S\ref{sec:spec}, the reason is that the wings of the
Ly$\alpha$ profile are comprised by photons originating in the very dense
regions of the galaxy having to diffuse far from the line center in order to
escape, and since this is also where the most of the dust is residing, these
photons have a larger probability of being absorbed.
For the two least massive galaxies, S108sc and S115sc, the ISM is neither very
dense nor very metal-rich, meaning that photons escape rather easily.
Consequently, the line is
not particularly broadened and most of the photons escape the galaxies.

%

\subsubsection{Extended surface brightness profile}
\label{sec:ext}

\begin{figure*}
\epsscale{1}
\plotone{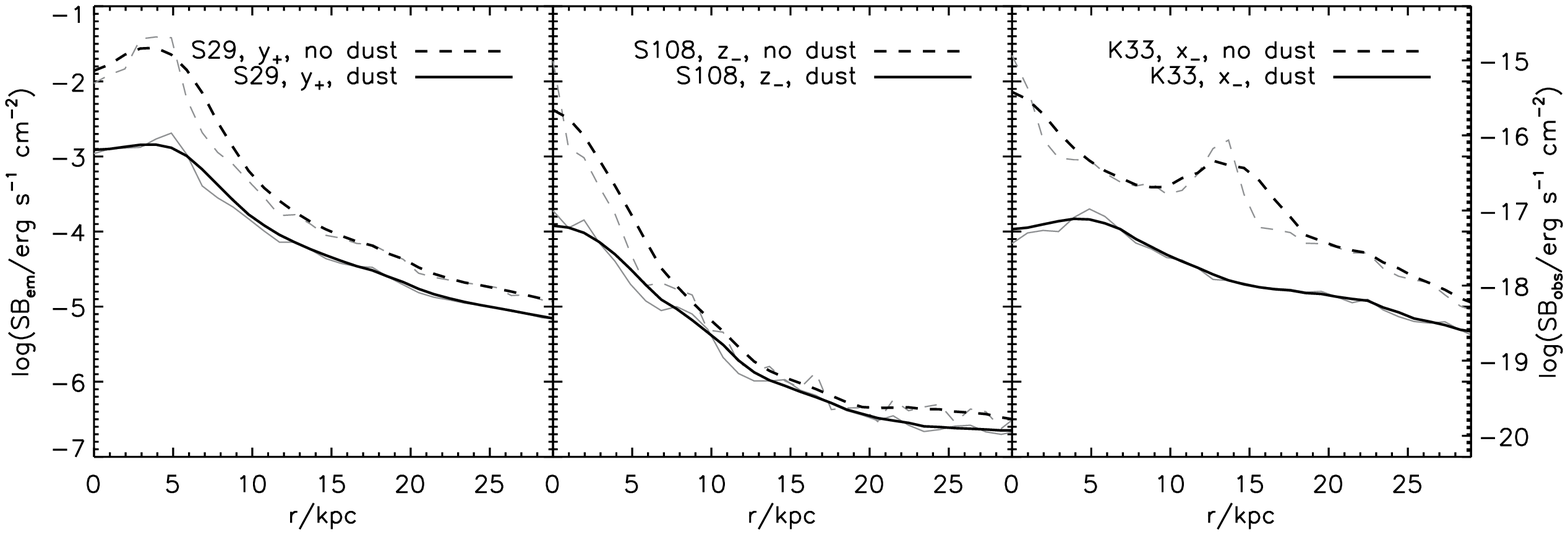}
\caption{Ly$\alpha$ surface brightness (SB) profiles of three of the simulated
         galaxies as observed from three arbitrary directions, with
         (\emph{solid}) and without (\emph{dashed}) dust.
         While the intrinsic SB profiles are shown in grey, the black lines
         show the profiles convolved with a Gaussian kernel corresponding to
         a seeing of $0\farcs5$.
         In general, the inclusion of dust tends to ``smooth out'' the
         profiles, effectively resulting in a more extended SB profile.}
\label{fig:SBpanel}
\end{figure*}

The fact that more photons are absorbed in the bright regions tend to
``smooth out'' the SB profiles of the galaxies.
Young galaxies have often been found to be more extended on the sky when
observed in Ly$\alpha$, compared to continuum band observation
\citep{mol98,fyn01,fyn03,nil07}.
\citet{lau07} found that, even without dust, resonant scattering itself may
cause an extended Ly$\alpha$ SB profile. Including dust merely adds to this
phenomenon, since steep parts of the profile are flattened.
Figure \ref{fig:SBpanel} shows this effect for three arbitrarily chosen
galaxies and directions.
Extended emission is seen also on larger scales (tens of kpc), although this
is more likely to be caused by cold accretion onto dark matter halos
\citep{nil06}.

%


\subsection{Temporal fluctuations}
\label{sec:t}

Many factors play a role in determining the exact value of $f_{\mathrm{esc}}$.
Although the metallicity, and hence the state of matureness of the galaxy, as
well of the size of the galaxy seem to be the most significant property
regulating $f_{\mathrm{esc}}$, less systematic factors like the specific
configuration of gas and stars are also likely to have a large influence.
The scatter in Fig.~\ref{fig:fesc_M} is probably due to this effect. To
get an idea of the fluctuations of $f_{\mathrm{esc}}$ with time, RT
calculations was run for snapshots of K15 from 100 Myr before ($z = 3.8$) to
100 Myr after ($z = 3.5$) the one already explored at $z = 3.6$.
In this relatively short time, neither $Z$ nor $M_{\mathrm{vir}}$
should not evolve much, and thus any change should be due to stochastic
scatter.

\begin{figure}
\epsscale{1}
\plotone{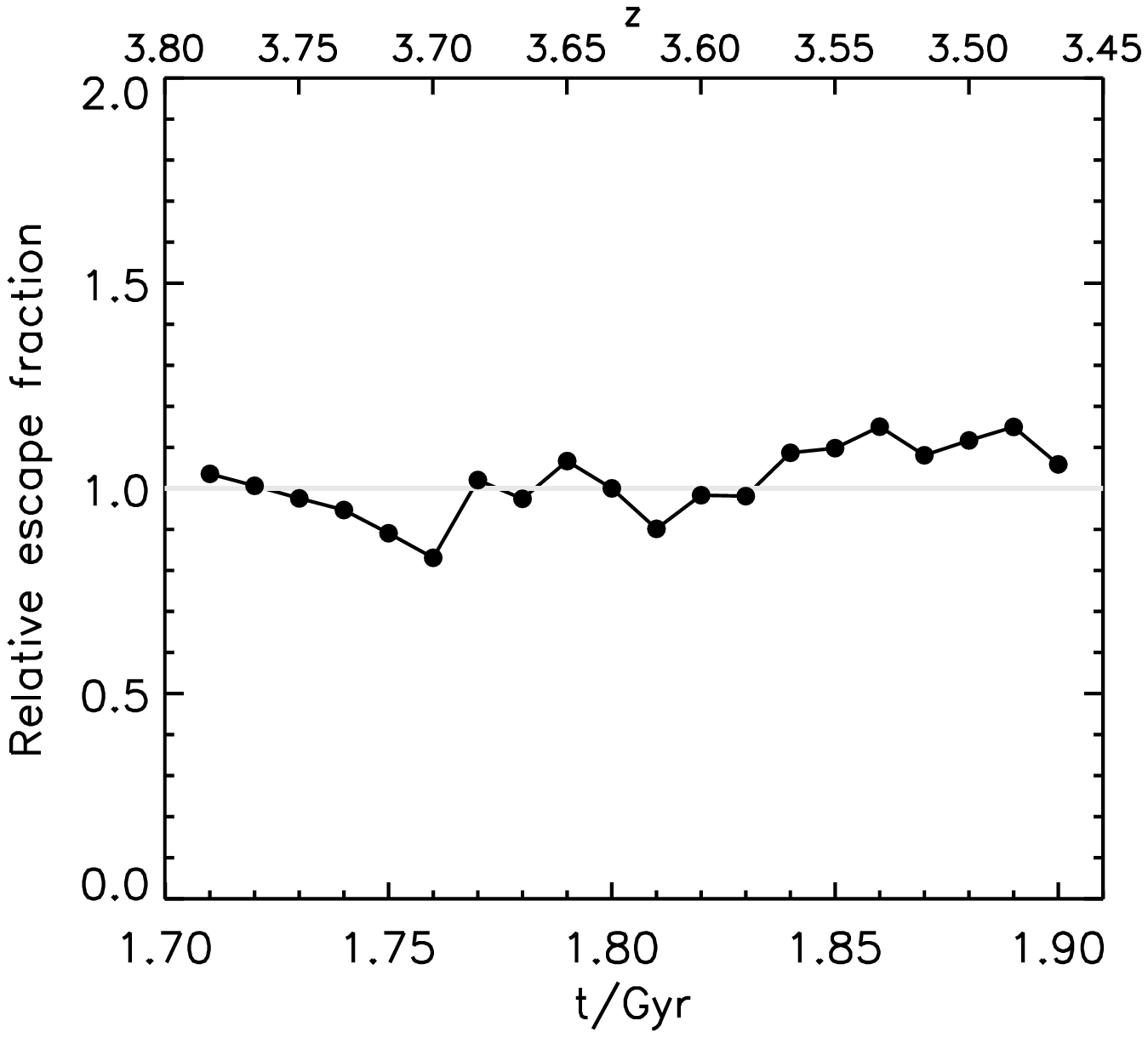}
\caption{Escape fraction $f_{\mathrm{esc}}$ from the galaxy K15 as a function
         of time $t$ over a period of 200 Myr, normalized to $f_{\mathrm{esc}}$
         at $t = 1.8$ Gyr. The dispersion
         of $f_{\mathrm{esc}}$ over time is quite
         small, indicating the the dispersion in Fig.~\ref{fig:fesc_M} is due
         to galaxy-to-galaxy variations rather than the escape fractions of the
         individual galaxies fluctuating.}
\label{fig:fesc_t}
\end{figure}

Figure \ref{fig:fesc_t} shows the variation of $f_{\mathrm{esc}}$ during this
time interval, demonstrating that
the temporal dispersion is of the order 10\% on a 10 Myr scale.
This suggests that the scatter seen in $f_{\mathrm{esc}}(M_{\mathrm{vir}})$
(Fig.~\ref{fig:fesc_M}) reflects galaxy-to-galaxy variations rather than
$f_{\mathrm{esc}}$ of the individual galaxies fluctuating strongly with time.

%


\section{Parameter study}
\label{sec:parstud}

\begin{figure}
\epsscale{1}
\plotone{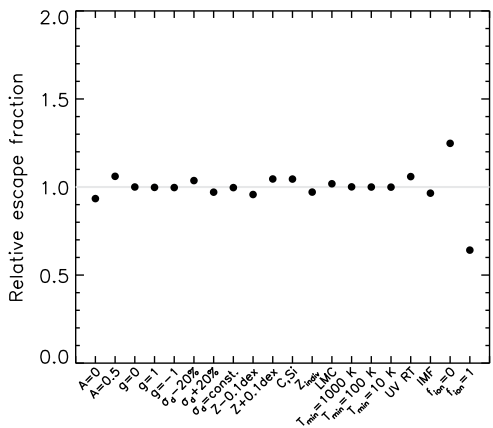}
\caption{Relative escape fraction $f_{\mathrm{esc}}$ from the galaxy K15 as a
         ``function'' of
         model, i.e.~simulation where all parameters but one are equal to that
         of the benchmark model used for the simulations in Sec.~\ref{sec:sim}.
         See \S\ref{sec:varpar} for an explanation of abcissa labels.
         Except for the factor $f_{\mathrm{ion}}$, the
         chosen benchmark model appears quite robust to varying other
         parameters.}
\label{fig:allpar}
\end{figure}

The adopted model of dust clearly involves a multitude of assumptions, some
more reasonable than others. To inspect the dependency of the outcome on the
values of the parameters a series of simulations of K15 was run,
varying the below discussed values. The resulting escape fractions compared to
that of the ``benchmark'' model, used for the simulations in
Sec.~\ref{sec:sim}, are shown in Fig.~\ref{fig:allpar}.

\subsection{Varied parameters}
\label{sec:varpar}

\emph{The albedo} $A$ of the dust grains. The chosen value of 0.32 is bracketed
by the values 0.5 and 0, i.e.~somewhat more reflective and completely
black, respectively. As expected, the higher the albedo, the higher the escape
fraction, but note that even completely black dust reduces $f_{\mathrm{esc}}$
by less than $10$\%. This is because the bulk of the
photons is absorbed in the very dense environments, where scattering off of
one grain in many cases just postpones the absorption to another grain.

\emph{The scattering asymmetry parameter} $g$. The three cases $g = 0,1,-1$
are tested, corresponding to isotropic scattering, total forward scattering,
and total backscattering. The difference from the benchmark model value of
$0.73$ is virtually
non-existent; the fact that most of the scatterings take place in the
dense environments makes the transfer be dominated by scattering on hydrogen.

\emph{The dust cross-section} $\sigma_{\mathrm{d}}$. \citet{fit07} showed that
the variance of the extinction curves (in the MW, normalized to $A_V$) is
approximately 20\% at the Ly$\alpha$ wavelength.
Decreasing (increasing) the dust cross-section by this quantity increases
(decreases) the escape fraction as
expected, but not by more than $\sim$$\pm5$\% percent.
Also, a constant cross-section
with $\sigma_{\mathrm{d}} = \sigma_{\mathrm{d}}|_{x=0}$ was tested, but with no
notable effect.

Likewise, a variance of \emph{the reference metallicity} $Z$ is present from
sightline to sightline in the Magellanic Clouds, probably at least 0.1 dex.
Using a smaller (larger) reference $Z$ makes the metallicity in the simulations
comparatively larger (smaller), with a larger (smaller) dust density as a
result and hence a smaller (larger) escape fraction. Since the escaping photons
represent different sightlines in the galaxies, it is fair to use the average
$Z$ (for the individual metals) in Eq.~(\ref{eq:nd}), but to investigate the
sensitivity on the reference metallicity, simulations with $Z_{\mathrm{SMC}}$
increased (reduced) by 0.1 dex was run, resulting in a $-5$\% ($+5$\%) change
in $f_{\mathrm{esc}}$.

Letting $n_{\mathrm{d}}$ scale with the metallicity of only C and Si instead
of the total metallicity does not alter $f_{\mathrm{esc}}$ much either. This
model is relevant since C and Si probably are the main constituents of dust.

Although the metallicity of the Magellanic Clouds is smaller than that of the
MW, the \emph{relative} abundances between the various elements are more or
less equal. Small deviations do exist, however, but letting $n_{\mathrm{d}}$
scale with the metallicity of the individual metals does not change the results
much (point labeled ``$Z_{\mathrm{indiv}}$'' in Fig.~\ref{fig:allpar}).


\emph{Dust type}. Using an LMC extinction curve instead of SMC results in a
slightly (few \%) larger escape fraction, since the quantity
$\tau_{\mathrm{d}} \propto
n_{\mathrm{d}} \sigma_{\mathrm{d}} \propto
\sigma_{\mathrm{d}}/Z_0$ is roughly 10\% lower for the LMC than for the SMC.

\emph{The minimum temperature} $T_{\mathrm{min}}$ of the simulations. The
cosmological simulation includes cooling of the gas to $\sim$$10^4$ K. Since
the temperature affects the RT of Ly$\alpha$, we artificially lowered the
temperature of the cells with  $T \simeq 10^4$ K  to $10^3$, $10^2$
(approximately the temperature of the cold neutral medium),
and $10$ K (approximately the
temperature of a molecular cloud), to see if not including sufficient cooling
could affect the results. However, as is seen in Fig.~\ref{fig:allpar}, the
difference is insignificant.

\emph{The ionizing UV RT scheme}. As mentioned already, implementing a more
realistic UV RT scheme does not alter the outcome of the non-dusty Ly$\alpha$
RT significantly. When including dust, as seen from Fig.~\ref{fig:allpar}
the improved RT results in a slightly increased $f_{\mathrm{esc}}$, although
less than 10\%. The reason
is that this RT is more efficient than the ``old'' RT scheme at ionizing the
neutral gas in the immediate vicinity of the stars and, accordingly, at
lowering the dust density. However, in these regions the gas density is so high
that ionization in most cases is followed by instantaneous recombination, and
hence the physical state of the gas in the case of the improved RT is not
altered significantly.

\emph{The initial mass function}.
Since a
Salpeter IMF is more top-heavy than a Kroupa IMF, i.e.~produces relatively
more massive stars per stellar mass, it also yields a higher metallicity
and hence a higher absorption by dust. However, the increased feedback from the
massive stars serves as to counteract star formation, and these
two effects more or less balances each other. The ratio of the
Kroupa-to-Salpeter feedback energy is 0.617, while for the yield the ratio
is 0.575
(for oxygen). The result, as is seen from Fig.~\ref{fig:allpar}, is only a
slightly smaller escape fraction.

\begin{figure}
\epsscale{1}
\plotone{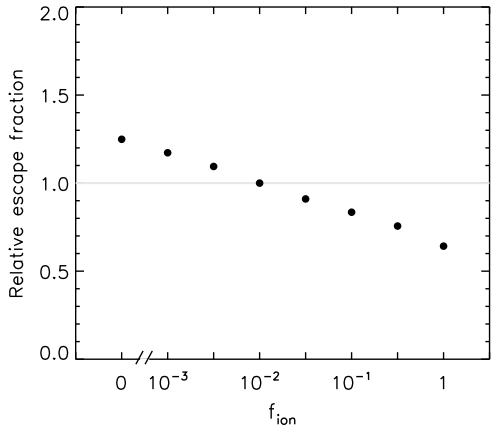}
\caption{Relative variation in escape fraction $f_{\mathrm{esc}}$ from the
         galaxy K15, as a function of
         $f_{\mathrm{ion}}$, the fraction of
         ionized hydrogen that contributes to the density of dust
         [Eq.~(\ref{eq:nd})]. Note the discontinuity on the abcissa axis.
         Even a little dust in the ionized region can
         affect $f_{\mathrm{esc}}$ quite a lot.}
\label{fig:subnosub}
\end{figure}

\emph{The fraction of ionized hydrogen}
$f_{\mathrm{ion}}$ contributing to the dust
density. Insufficient knowledge about the dust contents of ionized gas is by
far the greatest source of uncertainty in $f_{\mathrm{esc}}$, as is seen in
the figure. This effect is further investigated in Fig.~\ref{fig:subnosub},
where the relative escape fraction from K15 for different values of
$f_{\mathrm{ion}}$ is shown.

From the figure, it is seen that even a little amount of dust associated with
the ionized hydrogen can affect the escape fraction quite a lot. The reason is
that most of the scatterings and the absorption take place in the dense region
where also the star formation is high. In these regions supernova feedback
shockwaves will recurrently sweep through the ISM, heating and ionizing the
medium without significantly lowering the density.
For $f_{\mathrm{ion}} = 1$, the
calculated dust density of these regions is not affected, while
for $f_{\mathrm{ion}} = 0$, this effect
renders the gas virtually dustless, resulting in highly porous medium with
multiple possibilities for the photons of scattering their way out of the
dense regions.

The resulting escape fraction of the benchmark model lies approximately midway
between the two
extrema and if nothing else, $f_{\mathrm{esc}}$ can be regarded as having an
uncertainty given by the result of these extrema, i.e.~$\sim$20\%.
Nevertheless, we believe that $f_{\mathrm{ion}} = 0.01$ is a realistic value,
cf.~the discussion in \S~\ref{sec:iongas}.

\subsection{Resolution dependency}
\label{sec:resdep}

\begin{figure*}
\epsscale{1}
\plotone{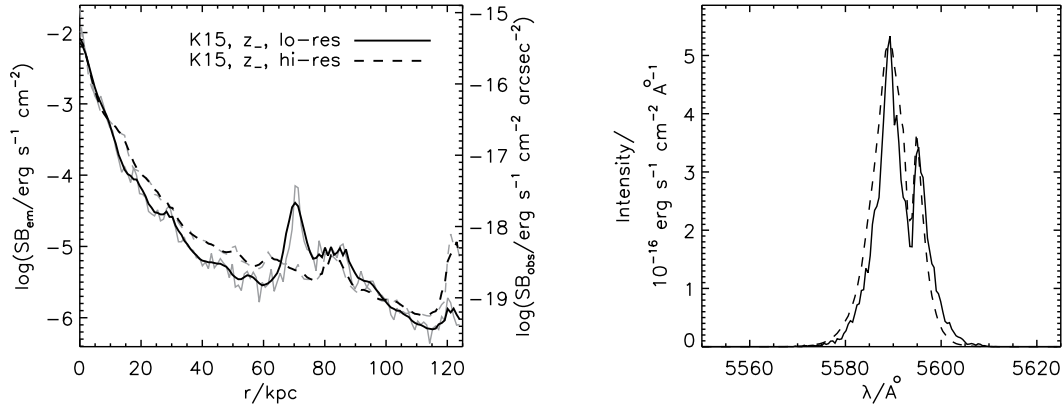}
\caption{Comparison of the surface brightness profile (\emph{left})
         and emergent spectrum (\emph{right}) of the radiation escaping the
         galaxy K15 when performed on galaxies simulated at low
         (\emph{solid}) and high (\emph{dashed}) resolution. Grey lines show
         the true profile and black lines show the profile convolved with a
         seeing of $0\farcs5$. While the SB in
         the central regions appear to agree nicely, the fact that we are
         comparing two different simulations make luminous regions in the
         outskirt appear somewhat shifted.}
\label{fig:lowres}
\end{figure*}

The galaxies studied have been extracted from a medium resolution cosmological
simulation and then re-simulated at 8 times higher resolution. To check if the
resolution is sufficient to trust the results, $f_{\mathrm{esc}}$-calculations
were also performed at the medium resolution. The resulting SB profile and
spectrum for K15 is seen in Fig.~\ref{fig:lowres}.
The resulting escape fraction is a tiny bit lower than for the hi-res
simulation, but less than 5\%.

To investigate the significance of the AMR structure, we also carried out
simulations progressively desolving the structure level by level. In K15, K33,
and S115, the
maximum level $\mathcal{L}$ of refinement is 7, where $\mathcal{L} = 0$
corresponds to the unrefined base grid of $128^3$ cells. Eight cells of
$\mathcal{L} = \ell$ are desolved to $\mathcal{L} = \ell - 1$ by taking the
average of the physical parameters. Since temperature reflects the internal
energy of a body of gas, and since the combined velocity is given by momentum,
$T$ and $\mathbf{v}_{\mathrm{bulk}}$ are weighted by mass.
\begin{figure}
\epsscale{1}
\plotone{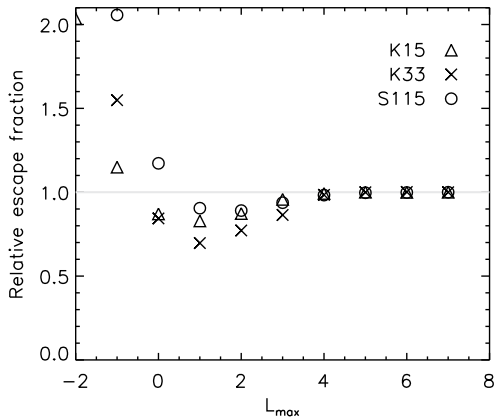}
\caption{Relative escape fractions $f_{\mathrm{esc}}$ from the three galaxies
         K15, K33, and S115 as a function of the
         maximum level $\mathcal{L}_{\mathrm{max}}$ of AMR refinement.
         $\mathcal{L}_{\mathrm{max}} = 0$ corresponds to having only the
         $128^3$ base grid, while $\mathcal{L}_{\mathrm{max}} < 0$ corresponds
         to desolving the base grid. For increasingly lower resolution,
         $f_{\mathrm{esc}}$ drops due to the low-density paths being smeared
         out with high-density regions. Eventually, however, when the
         resolution is so course that the central star- and gas-rich regions
         are mixed with the surrounding low-density region, $f_{\mathrm{esc}}$
         increases rapidly.} 
\label{fig:Lmax}
\end{figure}

Figure \ref{fig:Lmax} shows the resulting escape fractions of these three
galaxies. Desolving the first
few levels does not alter $f_{\mathrm{esc}}$ notably, indicating that the
galaxies are suffiently resolved. However, eventually we see the importance of
the AMR structure: with insufficient resolution, the clumpiness of the central,
luminous ISM is lost, ``smoothing out''
the low-density paths that facilitate escape,
and consequently $f_{\mathrm{esc}}$ decreases. When the resolution becomes even
worse, the central regions are averaged with the surrounding low-density gas,
so that most of the photons are being emitted from medium dense cells,
resulting in a small probability of scattering on neutral hydrogen, and hence
a small probability of being absorbed by dust.



\section{Summary and discussion}
\label{sec:sum}

We have undertaken numerical radiative transfer calculations for Ly$\alpha$
in high-redshift galaxies
including the effects of dust, with special emphasis on the fraction of photons
escaping the galaxies as well as the impact on the emergent spectra.
The abundance of, and the interaction probability with, dust is modeled by
scaling known extinction curves to the
(location-specific) metallicity of the galaxies, and destruction processes are
modeled by reducing the dust density in regions where hydrogen is ionized.

The main results are listed below:

\begin{itemize}

\item The escape fraction of Ly$\alpha$ seems to decrease with increasing
virial mass of the host galaxy. This is indicated by Fig.~\ref{fig:fesc_M},
which show that $f_{\mathrm{esc}}$ is close to unity for galaxies of
$M_{\mathrm{vir}} \sim 10^9$--$10^{10}$ $M_\odot$, while it falls off to a few
\% for $M_{\mathrm{vir}} \sim 10^{11}$--$10^{12}$ $M_\odot$.
This effect could be caused partly by the fact that smaller galaxies experience
less total star formation, so that the total amount of metals, and hence dust,
is smaller than for massive galaxies. However, since the same trend is seen for
the escape of ionizing UV radiation \citep{raz09}, for which dust is less
important, the chief factor is likely to be feedback energy of the smaller
galaxies being able to ``puff up'' and disorder its host to a larger degree
than for massive galaxies, due to their gravitational potential being smaller.

\item Although the cross-section of the dust is nearly independent of the
wavelength of light over the Ly$\alpha$ line, the spectrum is affected in a
highly wavelength dependent fashion:
close to the line center, the escape fraction is of the order 50\%, while it
quickly approaches zero in the wings. Consequently, the line is severely
narrowed, although its width may still reach several 100s of
km s$^{-1}$.

The reason is that different parts of the spectrum originates in physically
distinct environments of its host galaxy. The bulk of the Ly$\alpha$ photons
is produced in the central, dense regions, where hot stars ionize the
surrounding hydrogen which subsequently recombines, eventually resulting in
$\sim$2 Ly$\alpha$ photons for every 3 ionizing photon.
Since the optical depth of neutral hydrogen is so immense, in order to escape
the photons have to diffuse in frequency to either the
blue or the red side of the line center, where the cross-section falls off
rapidly. Hence, the wings of the spectrum emanates from the star-forming
regions, whereas the central parts to a high degree stems from the regions less
populated by stars, and from gravitational cooling of infalling gas.
Since the dust originates from stars, by far most of the dust is found where
the wings are produced, meaning that this part of the spectrum is affected the
most by the dust.

\item For the same reason, the surface brightness profiles of galaxies, when
observed in Ly$\alpha$, is ``smoothed out''. Without dust, the central parts
of the galaxies
are often very luminous compared to the outer parts, thus resulting in a
central bump in the SB profile. With dust, this bump is reduced, giving rise
to an even more ``flat'' SB profile, which can then be interpreted as an
extended SB profile when comparing to continuum bands.

\citet{lau07} found that even without dust, scattering effects alone can
explain extended Ly$\alpha$ profiles, although the very central parts may
exhibit a steeper profile. Including dust in the calculation tends to remove
these central bumps and thus flatten the profile.

\end{itemize}

The obtained results ($f_{\mathrm{esc}}$, SB, spectra) seem to be quite
insensitive to the assumed values of various parameters characterizing the
dust, such as dust albedo, scattering asymmetry, dust
cross-section, extinction curve, etc. Of the studied input parameters, the only
actual uncertainty comes from insufficient knowledge about the dust contents of
ionized gas, but this probably \emph{at most} introduces an error of
$\sim$20\%. This robustness against input parameters is convenient in the
sense that we can rely on the results, but is also a nuisance since
at least from Ly$\alpha$ observations, we should not expect to be able to
learn much about the physical properties of the dust itself.

As discussed in the introduction, previous attempts to determine Ly$\alpha$
escape fractions have been quite divergent, ranging from a few percent to
close to a hundred percent.
In this work we have shown that $f_{\mathrm{esc}}$ may indeed vary rather much
from galaxy to galaxy, and we propose a tentative evidence for a negative
correlation with galactic mass. Obviously, many factors play a role in
regulating $f_{\mathrm{esc}}$; in particular the age of a given galaxy will
be significant, since the dust accumulates over time.

Various authors have invoked different scenarios to explain their inferred
escape fractions, e.g.~galactic outflows, ionized paths, multi-phase medium,
and viewing angle.
We find no evidence that any single of these
scenarios should be dominating entirely the magnitude of $f_{\mathrm{esc}}$.
Rather, a mixture of gas kinematics, ISM clumpiness and ionization state, as
well as viewing angle will influence the
total observed Ly$\alpha$ luminosity, and hence the deduced escape fraction.
The investigated galaxies exhibit both outflows and infall, but not at
exceptionally high velocites, and artificially setting all velocites to zero
still allows plenty of radiation to escape (this was done for K15; the shape of
the spectrum is altered, but $f_{\mathrm{esc}}$ still is around 0.16).
Artificially erasing the clumpiness of the ISM decreases $f_{\mathrm{esc}}$ as
expected but also in this case much radiation still escapes.
The luminosity observed from different directions may vary by a factor of
$\sim$1.5 to $\sim$4. Note, however, that this factor could increase if the
effect of AGN was implemented in the cosmological simulations.

We have also investigated the variation in $f_{\mathrm{esc}}$ over a time span
of 200 Myr, and found that $f_{\mathrm{esc}}$ exhibits only minor fluctuations.

We have focused on the escape of Ly$\alpha$ photons from galaxies, and did
not take into account any absorption/scattering in the IGM. At $z = 3.6$, the
Universe is expected to be highly ionized, and thus most of the radiation
should be able to propagate freely once the host galaxy has been escaped.
Preliminary RT calculations using {\sc MoCaLaTA} on cosmological volumes
indicate that, on average, approximately 80\% of all of the radiation blueward
of the Ly$\alpha$ line center should be transmitted, while the Universe is
effectively transparent to radiation redward of the line center
(Laursen et al., in prep.).

\acknowledgments
We are grateful to Alex O. Razoumov for calculating the realistic temperature
and ionization state of the gas, and to Kim K.~Nilsson and Johan P.~U.~Fynbo
for giving useful comments on the draft.
The simulations were performed on the facilities provided by the Danish Center
for Scientific Computing.
The Dark Cosmology Centre is funded by the Danish National Research Foundation.
%

\appendix

\section{Acceleration schemes}
\label{sec:acc}

\subsection{Core-skipping scheme}
\label{sec:acc1}

For a photon with frequency close to the line center, the opacity of the gas is
so high that the photon does not diffuse significantly in space. In order to
accelerate to execution time of the code, these scatterings can be skipped by
drawing the velocity of the scattering atom not from a pure Gaussian but from
a Gaussian where the inner part is truncated, such that higher velocities and
hence large frequency shifts are
favored. This acceleration scheme is invoked whenever
$|x| \le x_{\mathrm{crit}}$, where
$x_{\mathrm{crit}}$ varies according to the physical conditions in the cell
\citep[for a detailed explanation, see ][]{lau09}.

With a dusty medium, we must investigate the possibility that the photon
would have been destroyed, had we \emph{not} used this acceleration scheme,
i.e.~the probability $P_{\mathrm{abs}}(x_{\mathrm{crit}})$ of absorption for a
photon initially in the core, before escaping the frequency interval
$[-x_{\mathrm{crit}},x_{\mathrm{crit}}]$.
Ultimately, we will determine this probability numerically, but to interpret
the result, we will first investigate the scenario analytically.
In the following calculation, factors of order unity will be omitted.

The probability per interaction that a photon with frequency $x$ be absorbed is
\begin{equation}
\label{eq:Pabsi}
p_{\mathrm{abs}}(x) = \frac{\tau_{\mathrm{a}}}
                           {\tau_{\mathrm{d}} + \tau_x}
                 \sim \frac{1}{1 + \phi(x)\tau_0/\tau_{\mathrm{a}}},
\end{equation}
since $\tau_{\mathrm{a}} \sim \tau_{\mathrm{d}}$ and
$\tau_x \sim \phi(x)\tau_0$.
Here, $\tau_{\mathrm{a,d,}x\mathrm{,0}}$ corresponds to the optical depth of this particular part of the
journey.
The number $dN(x)$ of scatterings taking place when the frequency of the
photon is close to $x$ is
\begin{equation}
\label{eq:dN}
dN(x) = N_{\mathrm{tot}} \phi(x) dx,
\end{equation}
where $N_{\mathrm{tot}}$ is the total number of scatterings before the
photon exits $[-x_{\mathrm{crit}},x_{\mathrm{crit}}]$,
i.e.~the total number of scatterings
skipped. Here we have assumed \emph{complete redistribution} of the frequency,
i.e.~there is no correlation between the frequency of the photon before and
after the scattering event. This is a fair approximation in the core
\citep{unn52b,jef60}. However, \citet{ost62} showed that once the photon
is in the wing, it has a tendency to stay there, only slowly drifting toward
the line center with a mean shift per scattering
$\langle \Delta x \rangle = -1/|x|$.

For the purpose of the current calculation, the Voigt profile is
approximated by a Gaussian in the core and a power law in the wing, such that
\begin{equation}
\label{eq:phiapp}
\phi(x) \sim \left\{ \begin{array}{ll}
e^{-x^2} & \textrm{ for } x < x_{\mathrm{cw}}\\
\frac{a}{x^2}           & \textrm{ for } x \ge x_{\mathrm{cw}},
\end{array}
\right.
\end{equation}
where $x_{\mathrm{cw}}$ marks the value of $x$ at the transition from core to
wing \citep[Eq.~(15) in][]{lau09}.

At each scattering, the probability of escaping the region confined by
$x_{\mathrm{crit}}$ is
\begin{eqnarray}
\label{eq:Pesc}
\nonumber
p_{\mathrm{esc}}(x_{\mathrm{crit}}) & = & 2 \int_{x_{\mathrm{crit}}}^\infty
                                          \phi(x) dx\\
& \sim & \left\{ \begin{array}{ll}
    \mathrm{erfc}\, x_{\mathrm{crit}} &
    \textrm{\, \, \, for } x_{\mathrm{crit}} < x_{\mathrm{cw}}\\
\frac{a}{x_{\mathrm{crit}}}   &
    \textrm{\, \, \, for } x_{\mathrm{crit}} \ge x_{\mathrm{cw}},
\end{array}
\right.
\end{eqnarray}
where erfc is the complimentary error function.

Using Eq.~(\ref{eq:dN}), the total probability of being absorbed can be
calculated as
\begin{equation}
\label{eq:Pabst}
P_{\mathrm{abs}}(x_{\mathrm{crit}}) = \int_0^{x_{\mathrm{crit}}}
                                       p_{\mathrm{abs}}(x) dN(x).
                   = N_{\mathrm{tot}} \int_0^{x_{\mathrm{crit}}}
                                       p_{\mathrm{abs}}(x) \phi(x) dx(x).
\end{equation}
The total number of scatterings before escape if the photon is not absorbed
is $N_{\mathrm{tot}} \sim 1/p_{\mathrm{esc}}$.
For $x_{\mathrm{crit}} < x_{\mathrm{cw}}$, from Eqs.~(\ref{eq:Pabsi}) and
(\ref{eq:Pesc}), Eq.~(\ref{eq:Pabst}) then evaluates to
\begin{equation}
\label{eq:Pabs1}
P_{\mathrm{abs}}(x_{\mathrm{crit}}) \sim \frac{1}
                                          {{\mathrm{erfc}} x_{\mathrm{crit}}}
   \int_0^{x_{\mathrm{crit}}}
   \frac{dx}{e^{x^2} + \tau_0/\tau_{\mathrm{a}}}.
\end{equation}
The exponential integral and the factor $1/\textrm{erfc}\, x_{\mathrm{crit}}$
is of the same order, but since the factor $\tau_0 / \tau_{\mathrm{a}}$ is of
the order $10^8 Z/Z_\odot$, Eq.~(\ref{eq:Pabs1}) will usually be negligible.

In the case of $x_{\mathrm{crit}} \ge x_{\mathrm{cw}}$,
\begin{equation}
\label{eq:Pabs2}
P_{\mathrm{abs}}(x_{\mathrm{crit}}) \sim
     \int_0^{x_{\mathrm{crit}}} \frac{dx}
     {a/\phi(x) + a\tau_0/\tau_{\mathrm{a}}}.
\end{equation}
This integral can be evaluated separately for the intervals
$[0,x_{\mathrm{cw}}[$ and $[x_{\mathrm{cw}},x_{\mathrm{crit}}]$. For the first,
the result is usually negligible, as in the case with Eq.~(\ref{eq:Pabs1}).
The second integral yields
\begin{equation}
\label{eq:Pabs22}
P_{\mathrm{abs}}(x_{\mathrm{crit}}) \sim \frac{1}{\mathfrak{t}}
(\tan^{-1}\frac{x_{\mathrm{crit}}}{\mathfrak{t}} -
 \tan^{-1}\frac{x_{\mathrm{cw}}}  {\mathfrak{t}}),
\end{equation}
where
\begin{equation}
\label{eq:a}
\mathfrak{t} \equiv (a\tau_0/\tau_{\mathrm{a}})^{1/2}.
\end{equation}

\begin{figure}
\epsscale{.55}
\plotone{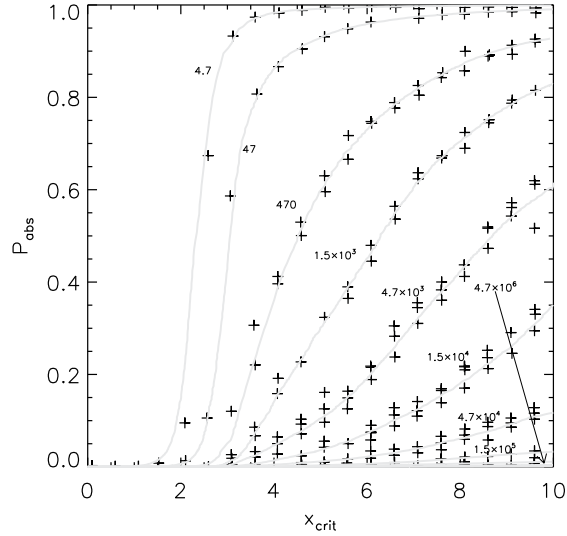}
\caption{Probability of absorption $P_{\mathrm{abs}}$ before escaping the
         region of the line confined by the value $x_{\mathrm{crit}}$, for a
         series of different values of $a\tau_0/\tau_{\mathrm{a}}$ (labeled
         at the corresponding lines).}
\label{fig:Pabs}
\end{figure}

For $x_{\mathrm{crit}} \ge x_{\mathrm{cw}}$, the assumption of complete
redistribution becomes very inaccurate, as the photon spends considerably
more time in the wings, with a larger probability of being destroyed. However,
Eqs.~(\ref{eq:Pabs22}) and (\ref{eq:a}) reveals a signature of the behavior of
$P_{\mathrm{abs}}(x_{\mathrm{crit}})$, viz. that it has a ``$\tan^{-1}$-ish''
shape, and that it scales not with the
individual parameters $a$, $\tau_0$, and $\tau_{\mathrm{a}}$, but with their
interrelationship as given by the parameter $\mathfrak{t}$. To know exactly the
probability of absorption,
a series of Monte Carlo simulations were carried out
for a grid of different temperatures, gas densities, and dust densities. The
results, which are stored as a look-up table, are shown in Fig.~\ref{fig:Pabs}.
Indeed, the same fit applies approximately to different $T$,
$n_{\textrm{{\scriptsize H}{\tiny \hspace{.1mm}I}}}$, and $n_{\mathrm{d}}$
giving equal values of $\mathfrak{t}$.

Whenever the acceleration scheme is applied, a bilinear interpolation over
$\log(\mathfrak{t})$ and $x_{\mathrm{crit}}$ determines the appropriate value
of $P_{\mathrm{abs}}$. A univariate is then
drawn and compared to $P_{\mathrm{abs}}$, thus
determining if the photon is absorbed or allowed to continue its journey.
Note that under most physically realistic conditions, only low values of
$P_{\mathrm{abs}}$ are actually met. However, when invoking the acceleration
scheme many times, the probability of absorption may become significant.

\subsection{Semi-analytical scheme}
\label{sec:acc2}

In very dense regions, defined in the simulations as cells with
$a\tau_0 > 2000$,
the code can be further accelerated by calculating analytically the
characteristics of a photon escaping such a region. \citet{lau09} found that
the spectrum of photons escaping a sufficiently dense cube is characterized
by the analytical solution for the spectrum from an equivalent ``slab'' of gas
found by \citet{har73}, with the exception that the independent variable
$a\tau_0$ (where $\tau_0$ is measured from the center to the face of the cube)
be replaced by $\eta a\tau_0$, where $\eta = 0.71$ is a fitting
parameter \citep[for a detailed explanation, see][]{lau09}.

With the inclusion of dust, we must calculate the possibility of the photon
being absorbed in such a cube. From the above, we might expect that replacing
$a\tau_0$ by $\eta a\tau_0$ and $\tau_{\mathrm{a}}$ by $\eta\tau_{\mathrm{a}}$
in the slab-relevant equation for the escape fraction [Eq.~(\ref{eq:neufesc})]
yields the relevant solution.
In fact, even better fits can be achieved by also replacing the square root
by an exponantiation to the power of 0.55. That is, every time the
semi-analytical acceleration scheme is invoked, a univariate is drawn and
compared to the quantity
\begin{equation}
\label{eq:cube}
f_{\mathrm{esc}} = \frac{1}
   {\cosh\left( \zeta'
         \left[ \eta^{4/3}
                (a\tau_0)^{1/3}
                (1-A)\tau_{\mathrm{d}}
         \right]^{0.55}
         \right)},
\end{equation}
determining whether or not the photon should continue its journey.

\begin{figure}
\epsscale{.5}
\plotone{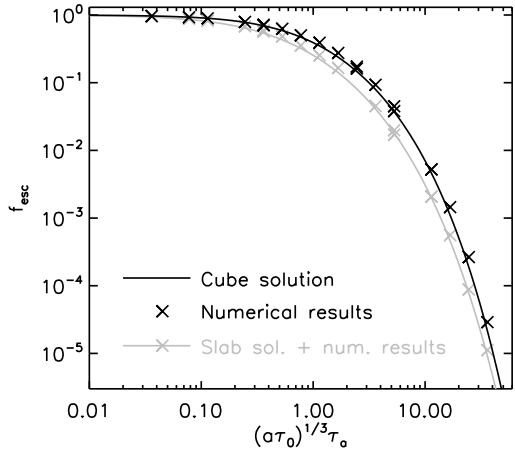}
\caption{Escape fractions (\emph{black crosses}) of photons emitted from the
         center of a cube of
         damping parameter $a$, line center optical depth $\tau_0$, and dust
         absorption optical depth $\tau_{\mathrm{a}}$, measured from the center
         to the face, compared to the analytical solution (solid black) given
         by Eq.~(\ref{eq:cube}). For comparison, the equivalent results for the
         slab (grey; the same as in Fig.~\ref{fig:neufesc}) are also displayed.}
\label{fig:cube}
\end{figure}

Figure \ref{fig:cube} shows the calculated escape fractions from a number of
cubes of different physical properties.
The tests performed in
\S\ref{sec:test} and many of the results of \S\ref{sec:sim} were performed both
with and without this acceleration scheme, all agreeing to a few percent within
statistical errors.

\subsection{Luminosity boosting scheme}
\label{sec:boost}

Since the vast majority of the photons are emitted within a relatively small
volume of the total computational domain, many photons are needed to reach good
statistics in the outer regions. In order to reach convergence faster, the
probability of emitting photons from low-luminosity cells can be artificially
boosted by some factor $1/w > 1$, later corrected for by letting the emitted
photon only
contribute with a weight $w$ to statistics (spectra, SB profiles, escape
fractions).

This factor is calculated for the $i$'th cell as
\begin{equation}
\label{eq:w}
w_i = 10^{\log(L_i/L_{\mathrm{max}})/b},
\end{equation}
where $L_i$ is the original luminosity of the cell, $L_{\mathrm{max}}$ is the
luminosity of the most luminous cell (not to be confused with
$\mathcal{L}_{\mathrm{max}}$), and $b$ is a ``boost buffer'' factor that
determines the magnitude of the boost; for $b = 1$, all cells will have an
equal probability of emitting a photon while for $b \to \infty$ the
probability approaches to original probability.

The optimal value of $b$ depends on the quantity and physical region of
interest. Since photons are absorbed primarily in the central regions,
$f_{\mathrm{esc}}$-calculation will usually converge fastest using
$b \to \infty$, and since after all most photons are recieved from this region
as well, the same counts for the spatially integrated spectrum. If one wishes
to investigate the SB or the spectrum of the outer regions, $b$ should not
simply be set equal to unity, however, since a significant fraction of the
photons received from here are photons originating in the central parts and
later being scattered in the direction of the observer. In this case, faster
convergence can be reached with $b \sim 1.5$ to a few tens.



\end{document}